\documentclass{article}
\usepackage{amsmath}
\usepackage{amssymb}
\usepackage{graphicx}
\usepackage{cite}
\usepackage{color}
\usepackage{todonotes}

\begin{document}
\title{Exploring complex networks via topological embedding on surfaces} %The complex structure of \\ maximal embedded graphs}
\author{
Tomaso Aste$^{1,2,3}$,
Ruggero Gramatica$^{4}$
and 
T. Di Matteo$^{4}$\\
\vspace{0.1cm} 
{\small
$^1$ School of Physical Sciences, University of Kent, CT2 7NZ, UK }\\
{\small $^2$ Applied Mathematics, Research School of Physics and Engineering,}\\ 
{\small The Australian National University, Canberra ACT 0200, Australia} \\
{\small $^3$ Department of Computer Science, University College London, }\\
{\small Gower Street - London - WC1E 6BT, UK }\\
{\small $^4$ Department of Mathematics,  King's College London, London, WC2R 2LS, UK} \\ 
{\small $\ast$ E-mail corresponding author: t.aste@ucl.ac.uk}
}

\maketitle

\begin{abstract}\noindent
We demonstrate that graphs embedded on surfaces are a powerful and practical tool  to generate, to characterize and to simulate networks with a broad range of properties.
Any network can be embedded on a surface with sufficiently high genus and therefore the study of topologically embedded graphs is non-restrictive.
We show that the local properties of the network are affected by the surface genus which determines the average degree, which influences the degree distribution and which controls the clustering coefficient. 
The global properties of the graph are also strongly affected by the surface genus which is constraining the degree of interwovenness, changing the scaling properties  of the network from large-world-kind (small genus) to small- and ultra-small-world-kind (large genus).
Two elementary moves allow the exploration of all networks embeddable on a given surface and naturally introduce a tool to develop a statistical mechanics description for these networks.
Within such a framework, we study the properties of topologically-embedded graphs which dynamically tend to lower their energy towards a ground state with a given reference degree distribution.
We show that the cooling dynamics between high to low 	`temperatures'  is strongly affected by the surface genus with the manifestation of a glassy-like transition occurring when the distance from the reference distribution is low. 
We prove, with examples, that  topologically embedded graphs can be build  in a way to contain arbitrary complex networks as subgraphs.
This method opens a new avenue to build geometrically embedded networks on hyperbolic manifolds.

\noindent
%\begin{keywords}
\end{abstract}

\section{Introduction}

In natural and artificial systems there exists a very broad variety of networks; 
indeed, networking is a crucial feature in information technologies, it is a vital skill in social behaviour and -more generally- it is at the base of the emergence of some of the fundamental properties of complex systems \cite{Barabasi99,Newman03,zhou2006,Calrarelli07,cohen2010complex}. 
Networks possess a wide range of properties and their structure can assume different forms depending on their function, their construction rules and their evolution dynamics.
From a general perspective, the structural properties of a network can be divided into two main categories: 
(i) the \emph{local structure}, concerning small portions of the graph which may vary from place to place and, typically, they are analyzed statistically (a well-known example is the degree distribution  \cite{Newman03}: distribution of the number of edges per vertex);
(ii) the \emph{global structure}, which concerns properties that involve the entire organization of the graph (a well-known example is the diameter  \cite{Newman03,cohen2010complex}: the longest shortest path between any two vertices).
It is understood that local properties and global properties are related, for instance, it has been shown that the degree distribution is affecting the diameter and, in particular, the presence of a few highly connected hub-vertices can reduce significantly the overall diameter  \cite{Newman03,cohen2010complex}.
However, the relation local/global is in general mediated through the hierarchical organization of the network and can result in non-trivial relations.
By their nature, local properties are easier to be measured and therefore, so far, they have attracted most of the attention in the literature. 

In this paper we show that by considering networks embedded on surfaces we can control an important global measure of complexity which is associated with the network \emph{interwovenness}.
It is quite intuitive that an increased interwovenness must be related with an increase of the complexity of the network structure, however, due to its global nature, a general quantification of this quantity is a very challenging task.
On the other hand, by considering the embedding of a network on a surface the interwovenness can be directly associated with the surface genus which is a non-negative integer number counting the number of handles in the surface \cite{Barth04}.
Let us recall that a network topologically embedded on a surface has the property that it can be drawn on the surface without edge-crossings (edges do not have to be straight).
The absence of edge-crossing is obviously a limitation on the degree of interwovenness of the network and it is therefore a constraint on its overall complexity. 
A sphere has no handles and genus $g=0$, a torus has one handle and $g=1$, a double torus has $g=2$, etc..
Intuitively, we can look at a handle in a surface as a `short-cut' that connects two distant parts. 
For instance, by joining the north and south poles of a sphere and pinching them together one can transform the sphere into a torus passing from $g=0$ to $g=1$.
It is clear that these short-cuts can be also used by the embedded network that in this way can link otherwise distant vertices. 
In this paper we consider only  simple graphs, where no more than one edge can directly connect two vertices and there are no loops connecting a vertex to itself.
If we take an orientable surface and we place $n$ vertices on it, we can then connect  with edges couples of vertices up to a point when no further edges can be inserted without generating edge-crossings. 
We call this graph \emph{maximal embedded graph} \cite{song2011nested,AsteSherr11} and, with the exemption of some special cases, it is a triangulation of the surface containing $n$ vertices and $3n+6(g-1)$ edges.
Let us note that some maximally embedded graphs on the Euclidean plane (genus equal to zero)  have been already proposed in the complex network literature \cite{Andrade05,Zhou05maxPlanar,gu2005simplex,song2012} and are known under the name of \emph{random Apollonian networks}.
By increasing the genus we can insert an increasingly larger number of edges in the maximal embedded graph.
Any other graph with $n$ vertices embedded on the same surface must be a subgraph of a maximal embedded graph (indeed, either it is a maximal embedded graph, or edges can be added to make it a maximal embedded graph). 
Furthermore, it has been proved that any graph can be embedded on an orientable surface with sufficiently large genus \cite{Ringel74}, therefore, by considering maximal embedded graphs we are embracing the whole family of all possible graphs.

\begin{figure}
\centering
\includegraphics[width=1\columnwidth]{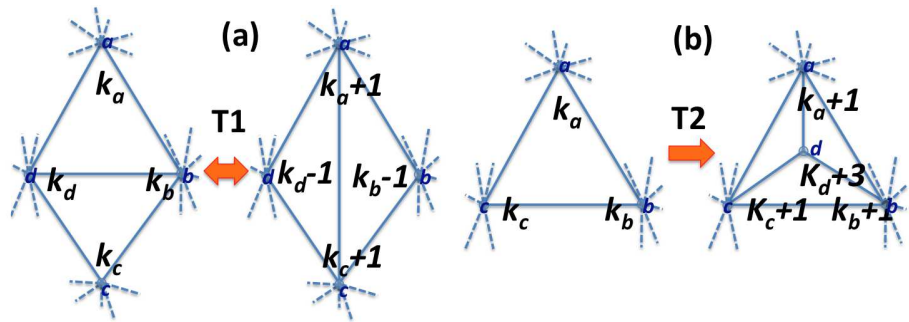}
\caption{\label{T1}
Exemplification of the two elementary moves which allow to explore all possible maximal embedded graphs on a given surface.
(a) The T1 move which consists in the switching of one edge between four vertices ($a$, $b$, $c$, and $d$);
(b) The T2 move, consisting in the insertion of a vertex ($d$) inside an existing triangle.
}
\end{figure}

One of the advantages of considering maximal embedded graphs is that there is a simple constructive way to build and modify them. In general, there exists a large number of models and construction methods that allow to build networks with desired properties and structures \cite{Newman03,cohen2010complex}.
Ideally, it would be desirable to be able to consistently generate networks with controlled and tunable properties both at local and global levels.
Such networks must be able to evolve and adapt following simple mechanisms and, eventually, allowing for a statistical mechanics kind of approach to be implemented.   
Maximal embedded graphs can be easily generated by starting from a seed embedded structure and then by letting them evolve by means of two elementary moves, known as T1 and T2  \cite{Alexander30,aste2Dfroth,Dubertret98}, which involve only  local changes and do not modify the embedding (see Fig.\ref{T1} and \cite{HyperTriang} for a toolbox to generate maximally embedded graphs and evolve them by means of these moves). 

It has been pointed out that complex networks naturally live in hyperbolic spaces  \cite{ADHhypnet04}.
There have been several studies \cite{boguna2010sustaining,Krioukov10,barthelemy2011spatial,daqing2011dimension} showing that relevant properties of complex networks can be mapped into properties of the hyperbolic geometrical space in which they are embedded.
An exact relation between space curvature and network  topology is provided by the discrete version of the Gauss-Bonnet formula  \cite{misner1973gravitation}.
When  the Gauss-Bonnet formula is applied to a triangulation of a surface it reveals that zero-curvature, Euclidean, space is associated with an average  degree equal to six, whereas negatively-curved hyperbolic spaces are tiled by triangulation with average degree larger than six and average degrees smaller than six are associated with elliptic positively-curved spaces.  
The curvature of the manifold is directly associated with the genus of the surface with the Euclidean space  having $g=1$, elliptic spaces having $g=0$ and hyperbolic spaces having $g>1$.
Any network can be embedded on a surface and for any network one can identify a surface-triangulation which contains the network as a subgraph.
Therefore, the study of the properties of maximally embedded graphs  give also insights on the properties of graphs geometrically embedded in spaces of arbitrary curvature.

The paper is organized as follows: in section \ref{s.1}, we introduce the idea of embedding networks on surfaces, we discuss the properties that embedded networks must obey, we describe the elementary moves T1 and T2, we define a simple energy function and we put the basis for a statistical mechanics description of these systems. 
In section \ref{s.2}, we perform an extensive numerical study of the properties of maximal embedded networks for the case of a regular reference degree distribution, discussing the effect of the surface genus and of the temperature on the structural and dynamical characteristics of these complex networks.
In section \ref{s.powLaw}, we investigate the case for a power-law  reference degree distribution.
In section \ref{s.embed}, we consider the problem of embedding an arbitrary network showing, through examples, that for any arbitrary complex network a surface triangulation which contains it as a subgraph can always be constructed. 
In section \ref{s.3}, conclusions and perspectives are given.

\section{Maximal Embedded Graphs}
%\section{Elementary moves, energy, dynamics and shortcuts: a statistical physics framework}
\label{s.1}

%In this paper we investigate triangulations on surfaces of different complexity where the surface complexity is quantified by its genus $g$ which measures the number of handles on the surface.
%For instance, a spherical surface has no handles and corresponds to $g=0$, a torus has one handle and it has $g=1$, etc..

\subsection{Topological properties}
As mentioned earlier, a maximal embedded graph on a surface of genus $g$ is a triangulation of the surface.
If $n$ is the number of vertices, $u$ the number of edges and $t$ the number of triangles, then the Euler's polyhedron formula \cite{Richeson08} provides us with a very simpe relation between these numbers
\begin{equation}\label{Euler}
n-u+t=2(1-g) \;\;.
\end{equation}
Furthermore, we have that,  in average, each vertex has $\left< k \right>$ incident edges and each edge has two vertices at its extremities (recall that we consider simple graphs that do not allow multiple edges between couples of vertices), yielding to
\begin{equation}
\left< k \right> n = 2u \;\;.
\end{equation}
Moreover, each triangle has three edges and each edge is in between two triangles, however some triangles may be self-neighbours  meaning that the edge may have the same triangle in both sides, implying
%If we further restrict our study only to triangulations where triangles cannot be self neighbours, then from the previous relation we can derive the average degree of the network because, in this case, each triangle has three edges and each edge is shared by two triangles and therefore $3t = 2u$.
%%Each triangle has three edges and each edge is shared by two triangles and therefore
%Instead, if we allow self-neighbouring triangels then an edge can also have the same triangle on both sides and the previous relation becomes more generically 
%\begin{equation}\label{incidence}
%3t =  \theta u \;\;, 
%\end{equation}
\begin{equation}\label{incidence}
0 < 3t \le 2 u \;\;, 
\end{equation}
where the upper bound $3t=2u$ is associated to the case when triangles cannot be self neighbours. % and  the limit $t \rightarrow 0$ is associated to the case when all triangles are self-neighbours.
By combining the previous two expressions with the Euler formula, Eq.\ref{Euler}, we get
\begin{equation}\label{av_n}
2 + 4 \frac{g-1}{n} <  \left< k \right>  \le 6 + 12 \frac{g-1}{n}\;\;,
\end{equation}
where the upper bound is achieved only when  triangles cannot be self-neighbours.

%consequently, the present simulations, we are investigating triangulations with a range of degrees between  $\left< k \right> \simeq 6$ ($g=0$) and  $\left< k \right> = 30$ ($g=2n+1$). 
%which indicates that in these networks the average degree is a topological invariant that is uniquely established by the genus and the number of vertices.
%This relation becomes an 
Incidentally, Eq.\ref{av_n} provides us with a lower bound for the genus required to embed any network.
Indeed, by construction, any network embedded on an orientable surface of genus $g$ must be a subgraph of a maximal embedded graph and therefore it must have an average degree smaller or equal than it.
Therefore, if $\left< k^* \right> $ is the average degree of an arbitrary network (not necessarily a triangulation), then from Eq.\ref{av_n} we have
\begin{equation}\label{low_g}
g \ge 1+  \frac{\left< k^* \right> - 6}{12} n\;\;.
\end{equation}
Although, this bound is achieved by maximally embedded graphs with no self-neighbouring triangles, it is  in general a rather loose bound for complex networks.
We can, for instance, note that for a large sparse network with in average less than six edges per vertex, the right-hand side of Eq.\ref{low_g} becomes negative, but $g$ must be a non-negative integer and therefore the bound becomes ineffective.
It must be noted that this bound can be also applied locally to any sub-graph.
This may give more restrictive limits especially when highly compact sub-structures are present.
For instance, if we have a $q$-clique (a configuration of $q$ vertices each-one connected with all the others), then for this sub-structure we can substitute $\left< k^* \right>  = q-1$ in Eq.\ref{low_g} obtaining $g \ge  (q-3)(q-4)/12$, which is the same bound proved by Ringel and Young in \cite{RingelPNAS68,Ringel74} for the solution of the map-coloring problem.
We can see that a tetrahedral clique ($q=4$) can be embedded on a sphere ($g \ge 0$), but a 5-clique ($K_5$, the complete graph with five vertices) requires already a larger genus $g \ge 1$.
This is in agreement with the Kuratowski theorem \cite{Kuratowski30} which proves that planar graphs cannot contain $K_5$ or $K_{3,3}$ (the complete bipartite graph with six vertices connected three by three) as minor (a \emph{minor} of  graph $G$ can be obtained from $G$ by edge deletion of edge contraction - the merging of two connected vertices).

\subsection{Elementary moves}
Maximal embedded graphs can be built by starting from a seed-graph and then by letting it evolve through simple elementary moves, called T1 and T2 \cite{Alexander30,aste2Dfroth,Dubertret98}.
The first move consists in the switching of an edge in a local configuration of four vertices  in which two second neighbour vertices become directly connected and vice-versa two first neighbour vertices become second neighbours.
This is shown in Fig.\ref{T1}(a).
New vertices can be added to the graph by means of the second move, T2,  which consists in the insertion of a vertex within an existing triangle generating in this way three new triangles, as shown in Fig.\ref{T1}(b) \cite{Alexander30}.
Similarly, vertices can be removed from the system by applying an inverse T2 move eliminating three triangles included inside a 3-clique.
These are local moves that do not change the global embedding of the graph. 
By means of T1 and T2 moves, it is easy to build and explore the entire class of maximal embedded graphs on a given surface \cite{Alexander30,aste2Dfroth}.
Let us note that some graphs generated by T2 moves only are known in the network-literature under the name of \emph{Apollonian networks}\cite{Andrade05,Zhou05maxPlanar,gu2005simplex,song2012}.
A toolkit to generate maximally embedded graphs on surfaces of arbitrary genera is available form Ref.~\cite{HyperTriang}.
This toolkit has been used to produce most of the numerical results presented in this paper.

\subsection{Energy}
We have now the tools to develop a statistical mechanics framework for maximal embedded graphs.
To this purpose it is convenient to introduce a  topological energy which can guide us in the exploration of configurations with different degree distributions.
Let consider a model (analogous to the configurational model \cite{Newman03}) where a  reference degree $k^*_i$ is given for each vertex.  
A simple form for the energy associates to the ground state ($E=0$)  the configuration where all vertices have the same degree as the reference network $k_i=k^*_i$ and a `cost' is instead associated to the deviations from reference degree: 
\begin{equation}
E = \sum_{i=1}^n (k_i - k^*_i)^2 \;\;\;.
\label{energy}
\end{equation}
This energy is a distance measure (the square of an Euclidean distance) from the ideal network with reference degree distribution $k^*_i$.
The simplest case is a regular reference degree distribution with all vertices having the same degree $k^*_i=\left<  k^* \right>$, this is the model first proposed in \cite{AsteSherr} and studied in several following works \cite{Davison00,Sherrington02,Kownacki04,Eckmann07,eckmann2012decay}. 
In this paper we will consider both the arbitrary distribution and the regular distribution cases.
%which associates the ground state ($E=0$) to the configuration with lowest topological disorder where all vertices have the same degree $k=\left< k \right>$ (if $\left< k \right>$ is integer).
%This energy associates a `cost' to the deviations from the average degree $\left< k \right>$ and it is therefore a distance measure (the square of an Euclidean distance) from an ideal regular network with degree $\left< k \right>$.

Our approach is to use a combined action of T1 and T2 moves to build a maximal embedded graph with a given number of vertices $n$ on a surface with given genus $g$ .
Once the graph is built we explore  achievable configurations by means of T1 moves only.

After a T1 move,  two vertices (e.g. $k_a$ and $k_c$ in Fig.\ref{T1}(a)) acquire a new edge and the other  two vertices (e.g. $k_b$ and $k_d$ in Fig.\ref{T1}(a)) loose one edge leaving the overall $ \left< k \right>$ unchanged.
As a consequence of such a move the energy change is:
\begin{eqnarray}\label{energyT1}
\Delta E &=& (k_a+1 - k^*_a)^2 + (k_c+1 -k^*_c)^2 + (k_b-1 - k^*_b)^2+ (k_d-1 - k^*_d)^2  \nonumber \\
&-& (k_a - k^*_a)^2 - (k_c- k^*_c)^2 - (k_b - k^*_b)^2 - (k_d -k^*_d)^2\;\;\;,
\end{eqnarray}
which is
\begin{equation}\label{energyT1b}
\Delta  E = 2(k_a - k^*_a) + 2(k_c-k^*_c) - 2(k_b-k^*_b) - 2(k_d-k^*_d) + 4 \;\;\;.
\end{equation}
Intriguingly, the energy variation as consequence of a T1 move is linearly dependent on the degrees of the four vertices involved in the T1 move.

%Intriguingly, the energy variation as consequence of a T1 move does not depend on $\left< k \right>$ (i.e. it is independent on the embedding) and it is linearly dependent on the degrees of the four vertices involved in the T1 move.
%One can immediately see that the T1 move decreases the system-energy iff $k_a+k_c < k_b+k_d$;
%meaning that the energy is decreased if the nodes with smaller degree acquire new connections at the expenses of the nodes with larger degree.
%We are therefore describing a local averaging process where inequalities in the degrees are gradually redistributed lowering in this way the total energy and driving the system towards the ground state.

\subsection{Dynamics}
A `temperature' $\beta^{-1}$ can be introduced and a statistical mechanics description can be implemented.
To this end, we adopt a Glauber--Kawasaki type of dynamics where a T1 move is performed accordingly with the probability \cite{AsteSherr}
\begin{equation}\label{prob}
\Pi(k_a,k_b,k_c,k_d) = \frac{1}{1+\exp(\beta \Delta E)}(1-\delta_{k_b,3})(1-\delta_{k_d,3})(1-\delta_{a,c})\;\;\;,
\end{equation}
where the Kronecker deltas enforce a `proper' triangulation. 
Specifically, the first two are preventing the vertex degree to become smaller than 3 and the third  avoids the formation of loops where edges connect a vertex to itself.

\section{Regular reference network, $k^*_i=\left<  k \right>$} \label{s.k*2} \label{s.2}

Let us first study the simplest case when the reference network is regular $k^*_i=\left<  k^* \right>$,
In this case  Eq.\ref{energyT1b} becomes:
\begin{equation}\label{energyT1c}
\Delta  E = 2(k_a + k_c - k_b - k_d) + 4 \;\;\;.
\end{equation}
revealing that the energy variation as consequence of a T1 move does not depend on $\left< k^* \right>$.
%This means that the local dynamic in itself is independent on the embedding (independent on the genus). 
One can also note that the T1 move decreases the system-energy if and only if  $k_a+k_c < k_b+k_d$; meaning that the energy is decreased if the nodes with smaller degree acquire new connections at the expenses of the nodes with larger degree that loose connections.
We are therefore describing a local averaging process where inequalities in the degrees are gradually redistributed lowering in this way the total energy and driving the system towards a regular network. % with equal degree at each vertex.

\begin{figure}
\centering
\includegraphics[width=0.7\columnwidth]{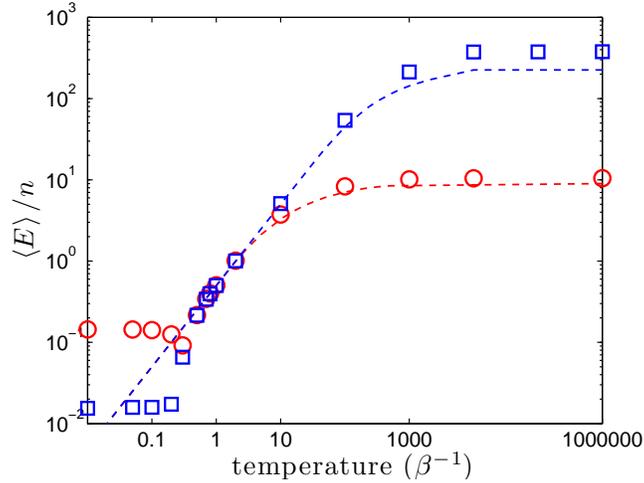}
\caption{\label{cool}
Average energy per vertex $\left< E \right>/n$ for the cooled states after $10^4\times n$  simulation steps at a given temperature $\beta^{-1}$ for maximal embedded graphs with $n=20000$.
Symbols:  $\square$ $g=n+1$; $\bigcirc$  $g=0$.
The dashed lines are the mean-field solutions associated with the two genera.
}
\end{figure}

\subsection{Mean-field solution}
The use of regular reference degree  introduces an important simplification of the form of the energy that becomes non-local and therefore suitable for a meanfield kind of approach.
The maximal embedded graphs that we are studying here have a very simple energy that depends only on the degree of the nodes. 
Nevertheless, interactions between vertices arise unavoidably from the network topology and its constraints, making any analytical study extremely challenging \cite{Tutte62,Brezin1978,boulatov86,Brezin90,Gross90,Godreche92}.
However, from a thermodynamic perspective we note that, at low enough temperatures, the energy should dominate over the entropy and it may be possible to describe the equilibrium state of the system by assuming all vertices as independent and reducing the topological correlations  to a simple constraint on the average degree. 
In such a \emph{meanfield approximation} the partition function, $Z_n$, factorizes and it can be written as:
%\subsubsection{Mean field solution}
\begin{equation}\label{mf}
Z_n \simeq \left( \sum_{k=3}^{n-1} \exp \left( -\beta \left( k- \left< k \right>\right)^2 + \lambda k \right) \right)^n\;\;\;;
\end{equation}
where the coefficient $\lambda$ is a Lagrange multiplier associated to the average degree and it is (implicitly) given by the condition:
\begin{equation}\label{mfK}
\left< k \right>=\frac{\partial}{\partial \lambda}  \ln Z_n \;\;\;.
\end{equation}
The average energy at equilibrium is given instead by
\begin{equation}\label{mfE}
\left< E \right>=-\frac{\partial}{\partial \beta} \ln Z_n \;\;\;,
\end{equation}
which can be calculated numerically. 
An analytical expression can be obtained in the limit of low temperatures revealing that the energy should decrease linearly with the temperature: $\left< E \right>/n \sim \beta^{-1}/2$.
In the opposite limit of high temperatures, we have instead that the energy is expected to tend to a constant: $\left< E \right>/n \simeq  (\left< k \right> - 3)^2$.

\subsection{Numerical results}%\label{s.2}
We have numerically generated maximal embedded graphs containing up to $n=20000$ vertices embedded on surfaces with different genera ranging between $g=0$ to $g=2n+1$.

\subsubsection{Random states} 
Let us start from high temperatures where the transition probability, Eq.\ref{prob}, becomes independent on the  degrees of the four nodes (except for the forbidden moves).
After a large number of moves the resulting equilibrium configurations become statistically stable and we shall refer to these configurations as the  \emph{random} states.
For $g=0$ some properties of this state are analytically known \cite{Tutte62,Brezin1978,boulatov86,Brezin90,Gross90,Godreche92}.
For instance, the degree distribution is:  $p(k) =  16(3/16)^k (k-2) (2k-2)!/[k!(k-1)!] $, which, in the tail region, is well described by an exponential law. 
This is in a satisfactory agreement with the mean field description that implies $p(k) =  ( \left< k \right> - 3 )^{k-3} ( \left< k \right> - 2)^{2-k}$.
However, when the genus increases, the distribution deviates from the exponential acquiring a power law behaviour in the region $k < \left< k \right> $ and simultaneously assuming a faster than exponential decay in the region of large degrees $k > \left< k \right> $.
We previously reported in \cite{ADHhypnet04} that the degree distribution for arbitrary genus appears to be well described by the following functional form $p(k) \propto  k^{-\xi} \exp(-\lambda k^{\gamma})$ where the parameters depend on the genus with $\xi$ passing from zero at $g=0$ to about 1 at $g=n+1$, $\lambda$ changing from about $0.3$  at $g=0$ to about $0.03$ at $g=n+1$ and $\gamma$ passing from about 1  at $g=0$ to about 2 at $g=n+1$ \cite{AsteSherr11}.
These changes of the parameters reflect the change in the degree distribution from exponential to power-law with a faster than exponential cut-off at large $k$.
A detailed analysis of the properties of this state is reported in \cite{AsteSherr11}.

\subsubsection{Equilibrium at finite temperatures}

In Fig.\ref{cool} we report the average energy measured after $10^4\times n$ time-steps at a given temperature  $\beta^{-1}$ on maximal embedded graphs with $n=20000$ vertices.
These values are averages over 10 cooling loops and the data refer to two different embeddings respectively with $g=0$ and $g=n+1$.
The dashed lines are the  mean-field predictions from Eq.\ref{mfE} (calculated in the limit $n \rightarrow \infty$ and by substituting the sum with an integral). % and the symbols are instead the measured values.
One can note that there is a range of temperatures in which the average energy follows very well the mean field prediction.
This is the regime where the network is at thermal equilibrium and the quadratic energy is dominating the degree distribution which is therefore well approximated by a normal distribution.
We can observe that at high temperatures ($\beta^{-1} \gg 1$),  the average energy saturates to a plateau following a qualitatively similar behaviour to the mean field prediction but with significant quantitative deviations. 
Indeed, at high temperatures, the statistical properties of the network are dominated by the entropic term which is associated to the number of triangulations that can be built on a surface of a given genus \cite{Tutte62}.
On the other hand, the mean field equations take into account the topological properties of the network only by constraining the average degree, and this turns out to be not sufficient for an accurate quantitative estimation.
%We can see from Fig.\ref{cool} that indeed the observed energies deviate from the mean field predictions at high temperatures.

\subsubsection{Freezing dynamics}
%Fig.\ref{cool} shows the residual average energy of the cooled stated after $10^4\times n$ attempted T1 moves at a given temperature  $\beta^{-1}$.
Starting from high temperatures ($\beta^{-1} \gg 1$) we can observe from Fig.\ref{cool}  that the residual energy decreases with the temperature indicating that --as expected--  states at lower temperatures are less disordered. 
As mentioned earlier, we observe that at intermediate temperatures ($\beta^{-1} \sim 1$) the system follows the mean field prediction, however, around  $\beta^{-1}\simeq0.5$ a change in the behavior occurs.
At low temperatures, the residual energy does not longer decrease with the temperature approaching instead a constant plateau.
This is the signature of a  `freezing' phenomena \cite{AsteSherr,Davison00,Sherrington02,Kownacki04,Eckmann07} typical of glasses \cite{AsteSherr,angell00,Cavagna09}.
% occurring in the in the cooling dynamics: at low temperatures the system is not any longer able to reach the equilibrium state in a finite time and the system becomes trapped into a frozen state.
%By comparing the mean-filed prediction for the average energy at equilibrium (Eq.\ref{mfE}) with the observed values we can see that 

\subsubsection{Frozen states}
%When the temperature is lowered the system becomes gradually less disordered.
At low temperatures ($\beta^{-1} \ll 1$) the maximal embedded graph should approach the ground state where all vertices have degree equal to $\left< k \right>$ (if it is an integer number).
However, we observe that the system cannot reach the ground state in any finite time.
This freezing phenomena was first observed in studies performed on planar triangulations with periodic boundary conditions ($g=1$) \cite{AsteSherr,Davison00,Sherrington02,Kownacki04,Eckmann07} and here it is retrieved in the general case of maximal embedded graphs on surfaces of arbitrary genus.
%it  has been observed in a study performed on embeddings on $g=1$ surfaces (e.g. the plane with  periodic boundaries conditions) that the system cannot reach the ground state in any finite time  \cite{AsteSherr,Kownacki04,Eckmann07}. 
Interestingly, we observe that  the complexity of the embedding surface (e.g. its genus $g$) is affecting the freezing dynamics and consequently the properties of the asymptotic states \cite{AsteSherr11}. 
In particular, triangulations with large genus start from random states with higher energies with respect to the low genus counterparts, but they cool faster and, within the same number of time-steps, they can reach frozen states with lower residual energy.
When the average degree is an integer, the degree distribution of the frozen states is characterized by a large fraction of vertices at a degree equal to the average and by two small fractions of `defective' vertices respectively with degree one above and one below the average (see \cite{AsteSherr11}).
A similar distribution is observed when the average degree is not an integer but it is still close to a natural number.
On the other hand, a different statistics and a different  dynamics are observed when  $\left< k \right>$ is near to an half-integer, an effect due to topological frustration.
 
\begin{figure}
\centering
\includegraphics[width=0.7\columnwidth]{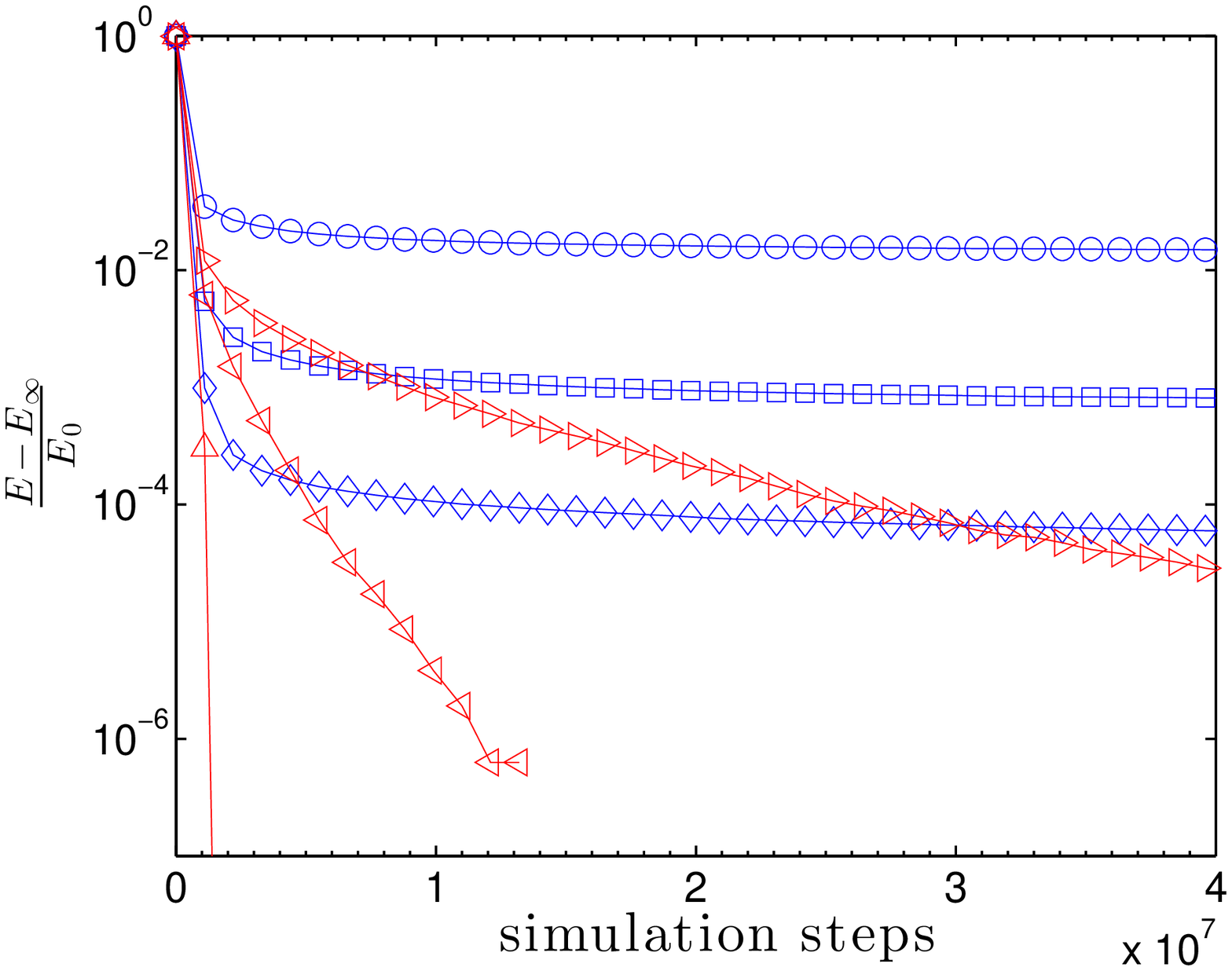}
\caption{\label{FrustratedCool}
Relative differences between the energy ($E$) and the expected energy of the ground state ($E_\infty$) divided by the average energy of the random state ($E_0$) as function of the number of simulation steps.
Symbols:
 $\bigcirc$ $g=1$,  $\left< k \right>=6$; 
$\square$ $g=10000$,  $\left< k \right>\simeq12$; 
$\diamond$ $g=20001$,  $\left< k \right>=18$;
$\triangleright$ $g=834$,  $\left< k \right>\simeq6.5$;
$\triangledown$ $g=12501$,  $\left< k \right>=7.5$;
$\triangle$ $g=17501$,  $\left< k \right>=16.5$.
}
\end{figure}

\subsubsection{Topological frustration}
When $\left< k \right>$ is in the middle between the two integers, $k_1$ and $k_2$, then the energy is minimized by a state with half of the vertices with degree equal to $k_1$ and  the other half  with degree equal to $k_2$.
This naturally introduces \emph{frustration} in the system  preventing the formation of a regular ground state.
In particular, the ground state energy passes from $E_\infty = 0$ for  $\left< k \right>$  integer to  $E_\infty = 0.25\,n$ for $\left< k \right>$ half integer. 
Intriguingly, we observe that this also strongly affects the dynamics.
In Fig.\ref{FrustratedCool} we compare the cooling dynamics in three networks with nearly integer average degrees (namely $g=1,\,10,000,\,20,001$ which correspond respectively to  $\left< k \right>\simeq 6,\, 12,\,18$) and in three other networks with nearly half-integer  average degrees (namely $g=834,\,g=2,501,\,g=17,501$, corresponding to  $\left< k \right>\simeq 6.5,\, 7.5,\,16.5$).
In all cases, the system is prepared in the random state at infinite temperature ($\beta=0$) and then the temperature is quenched to zero.
The figure reports the difference between the values of the energy  during the cooling dynamics after quenching and the expected energy of the ground state ($E_\infty$) divided by the average energy of the random state ($E_0$).
In this way, all plots start from 1 and they must approach 0 if the system is reaching equilibrium.
We observe that the unfrustrated systems with integer $\left< k \right>$  freeze to a plateau energy which decreases with the genus but it is well above $E_0=0$ for all the studied cases.
We instead observe that  frustrated systems with half-integer $\left< k \right>$ exhibit a fast dynamics to $E_\infty =0.25\,n$ which is reached in finite times with exponentially fast relaxation times.

\begin{figure}
\centering
\includegraphics[width=0.8\columnwidth]{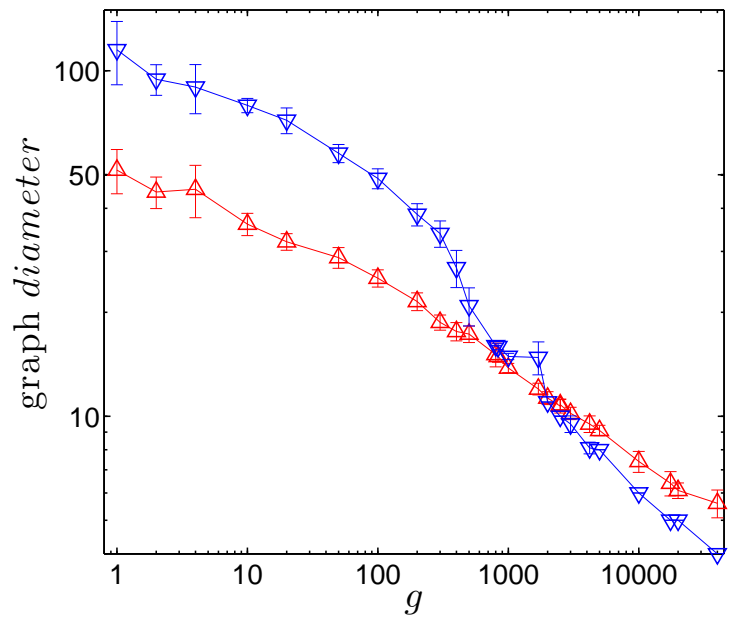}
\caption{\label{DiameterG}
Variation of the graph-diameter vs. the genus. 
Symbols: $\triangle$ random states;  $\triangledown$ frozen states.
The symbols correspond to averages over 10 independent constructions, the error bars are the standard deviations.
}
\end{figure}

\subsubsection{Shortcuts}
Let us now investigate the effect of the surface-genus on the global properties of maximal embedded graphs. 
As mentioned earlier, the presence of an handle in the surface provides a possible pathway for `shortcuts' between two distant parts of the network that otherwise could not be directly connected. 
It is intuitive to understand that by increasing the genus one also increase the number of shortcuts that can be inserted. 
Consequently the diameter of the graph (maximum distance between any two nodes)  must proportionally decrease when the genus increase.
However, we must stress that we are investigating a stochastic system that is spontaneously evolving through random weighted moves and therefore the relation between network-diameter and surface-genus is not straightforward.
In Fig.\ref{DiameterG} we report the graph diameters for samples with $n=20000$ vertices embedded on surfaces with different genus.
As expected, we can see that the diameter decreases with the genus.
We observe that this happens both for networks in the random and in the frozen states.
At low genus the random state has a smaller diameter than the frozen state, a feature that may be associated with the formation of hubs in the random configuration.
Intriguingly, we can see that at high genus the relation is reversed and frozen states have smaller diameters than random states, this may be associated to the formation of  branches in the random state.
Very similar results are retrieved by considering the average distance between vertices instead of the diameter.

\begin{figure}
\centering
\begin{tabular}{cc}
\includegraphics[width=0.5\columnwidth]{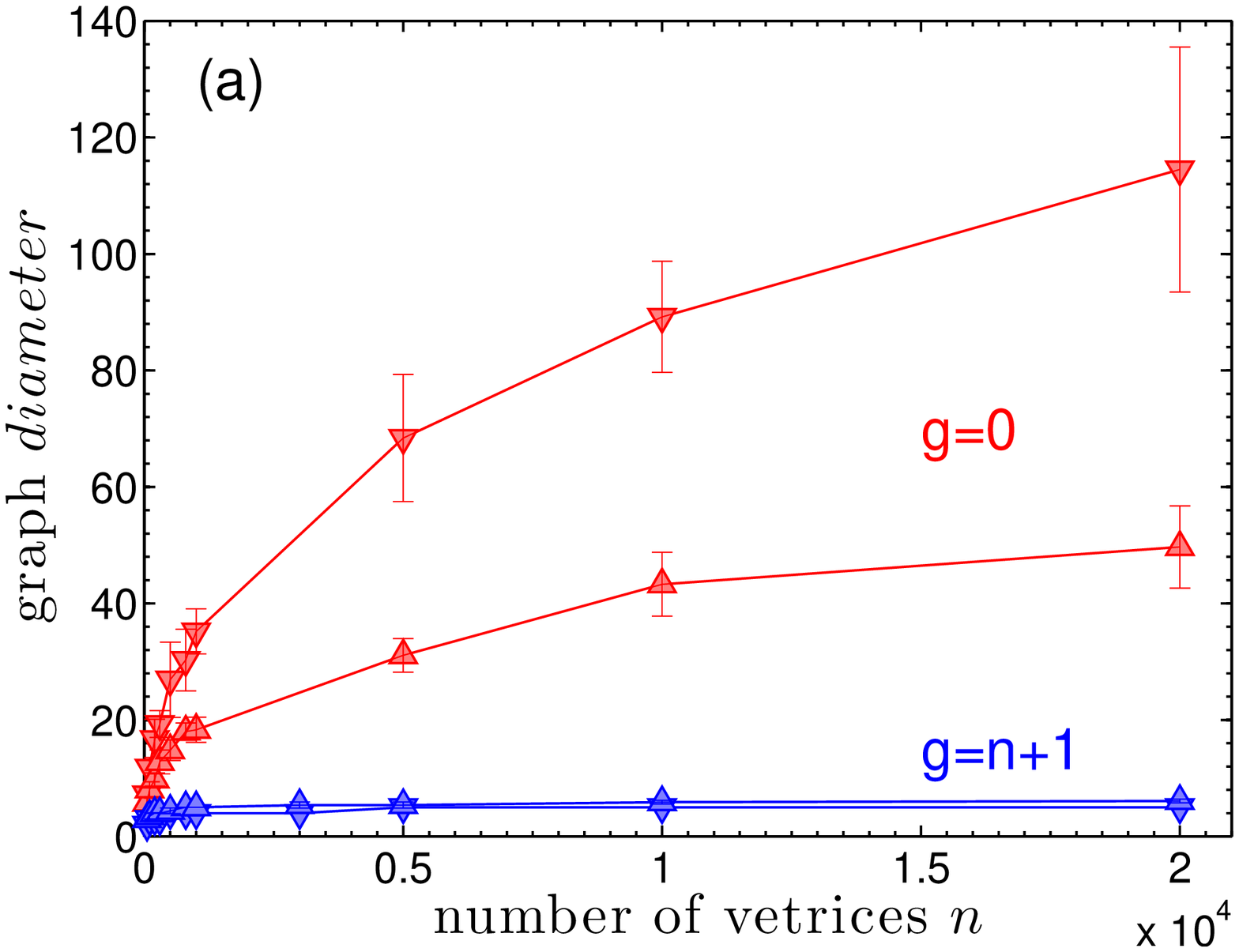}
&\includegraphics[width=0.5\columnwidth]{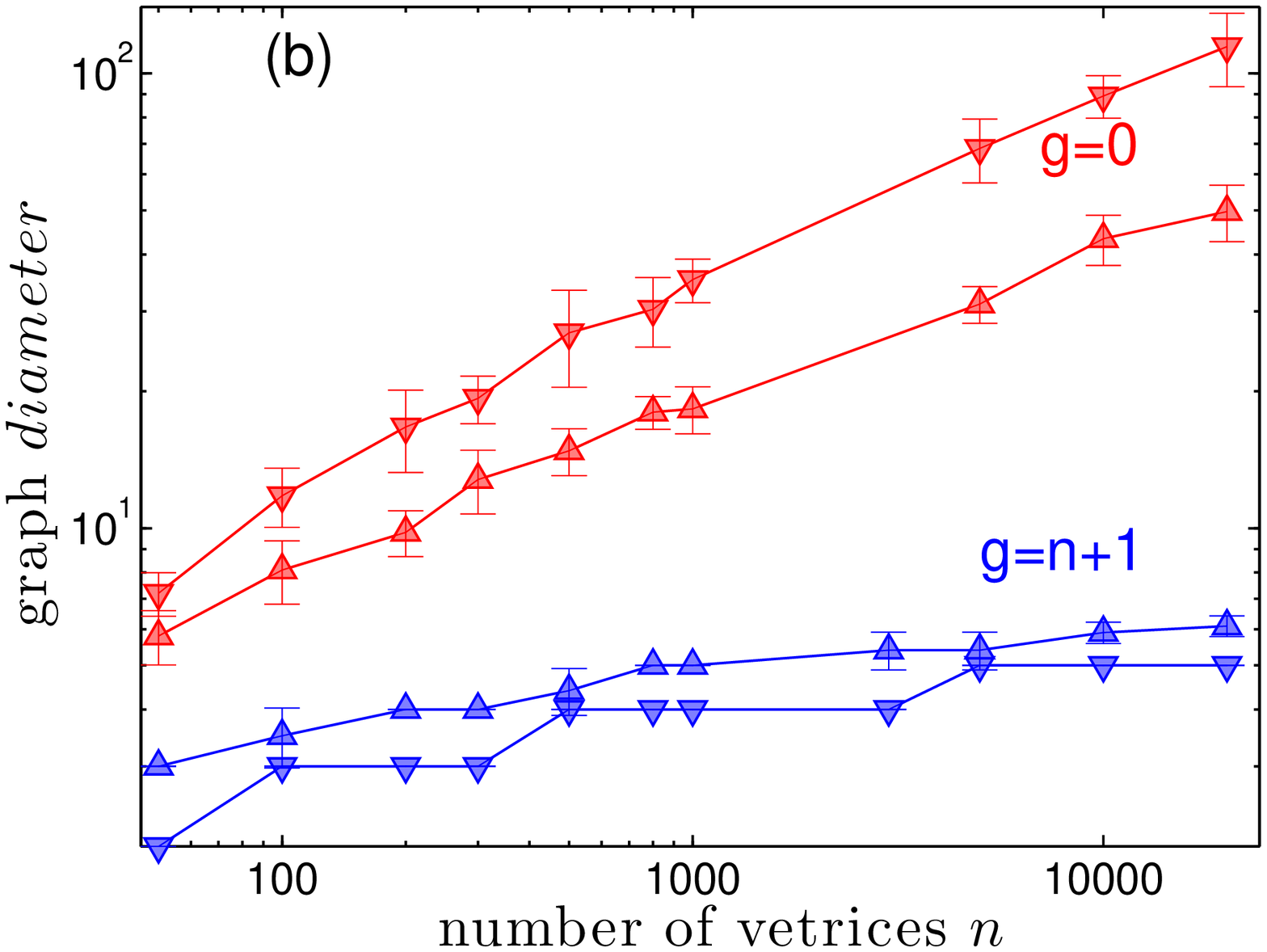}
\end{tabular}
\caption{\label{DiameterV}
Variation of the graph-diameter vs. the number of vertices. 
Symbols: $\triangle$ random states;  $\triangledown$ frozen states.
The symbols correspond to averages over 10 independent constructions, the error bars are the standard deviations.
(a) Plot with linear-linear axis.
(b) Plot with log-log axis.
}
\end{figure}
\begin{figure}
\centering
\begin{tabular}{cc}
\includegraphics[width=0.5\columnwidth]{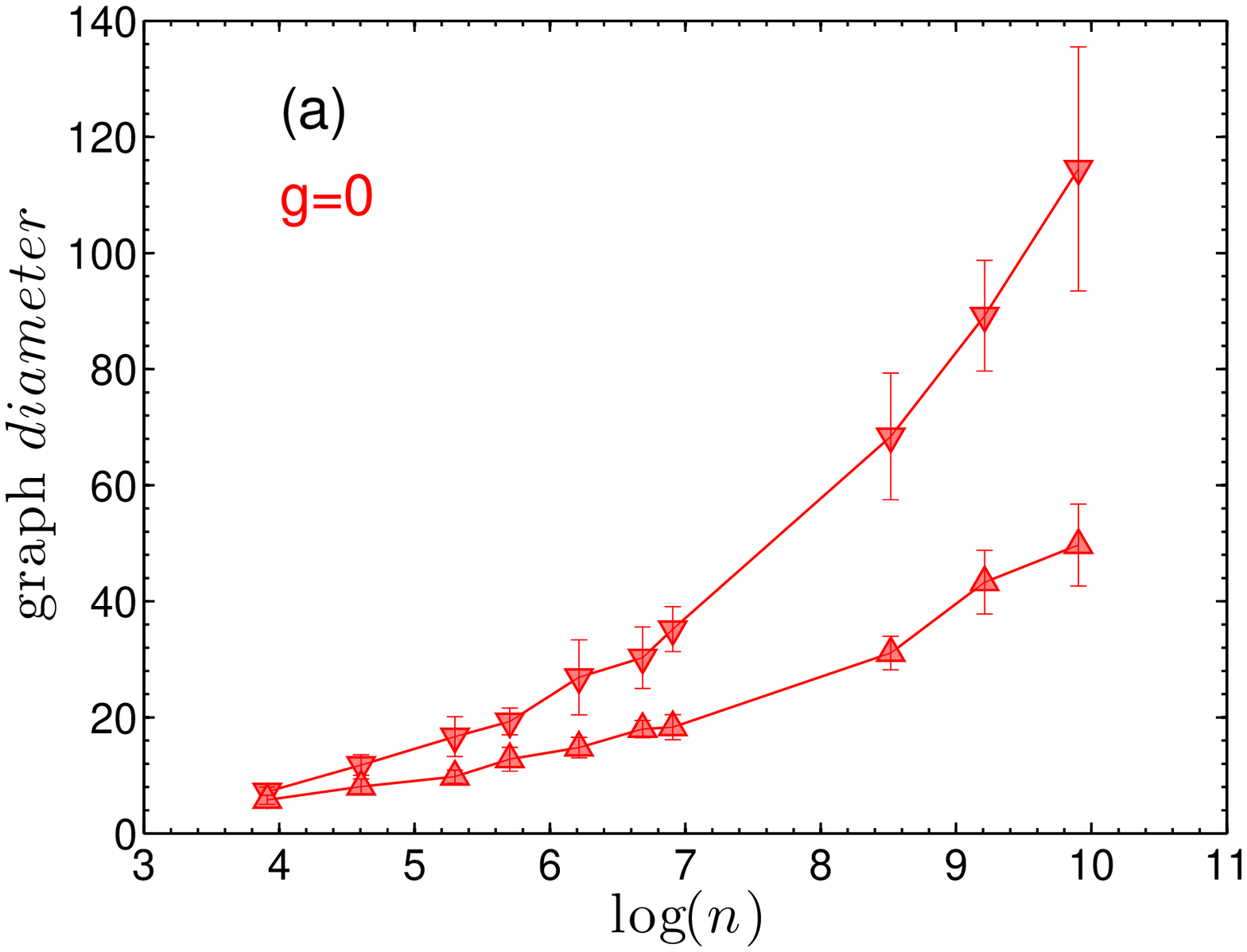}
&\includegraphics[width=0.5\columnwidth]{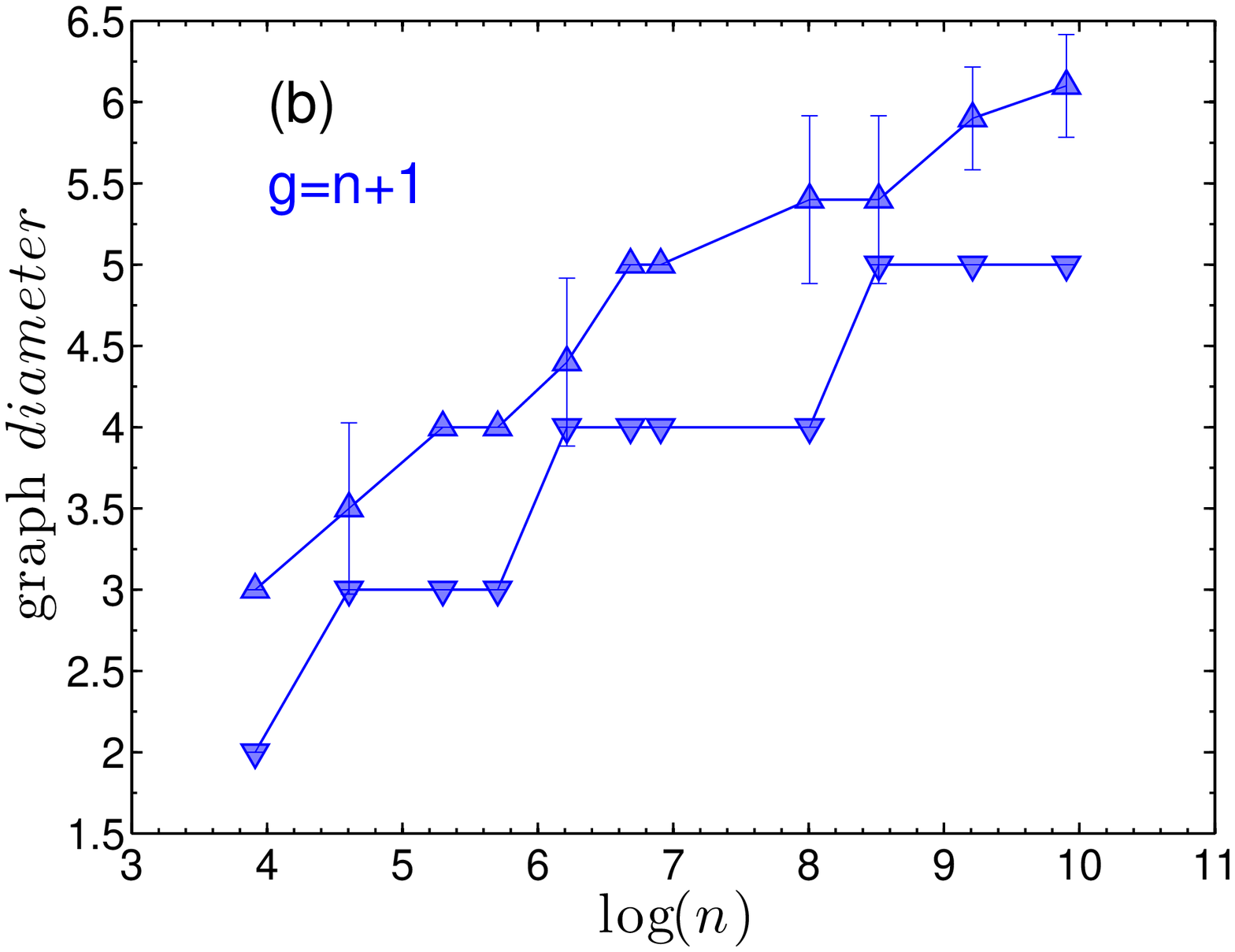}\\
\includegraphics[width=0.5\columnwidth]{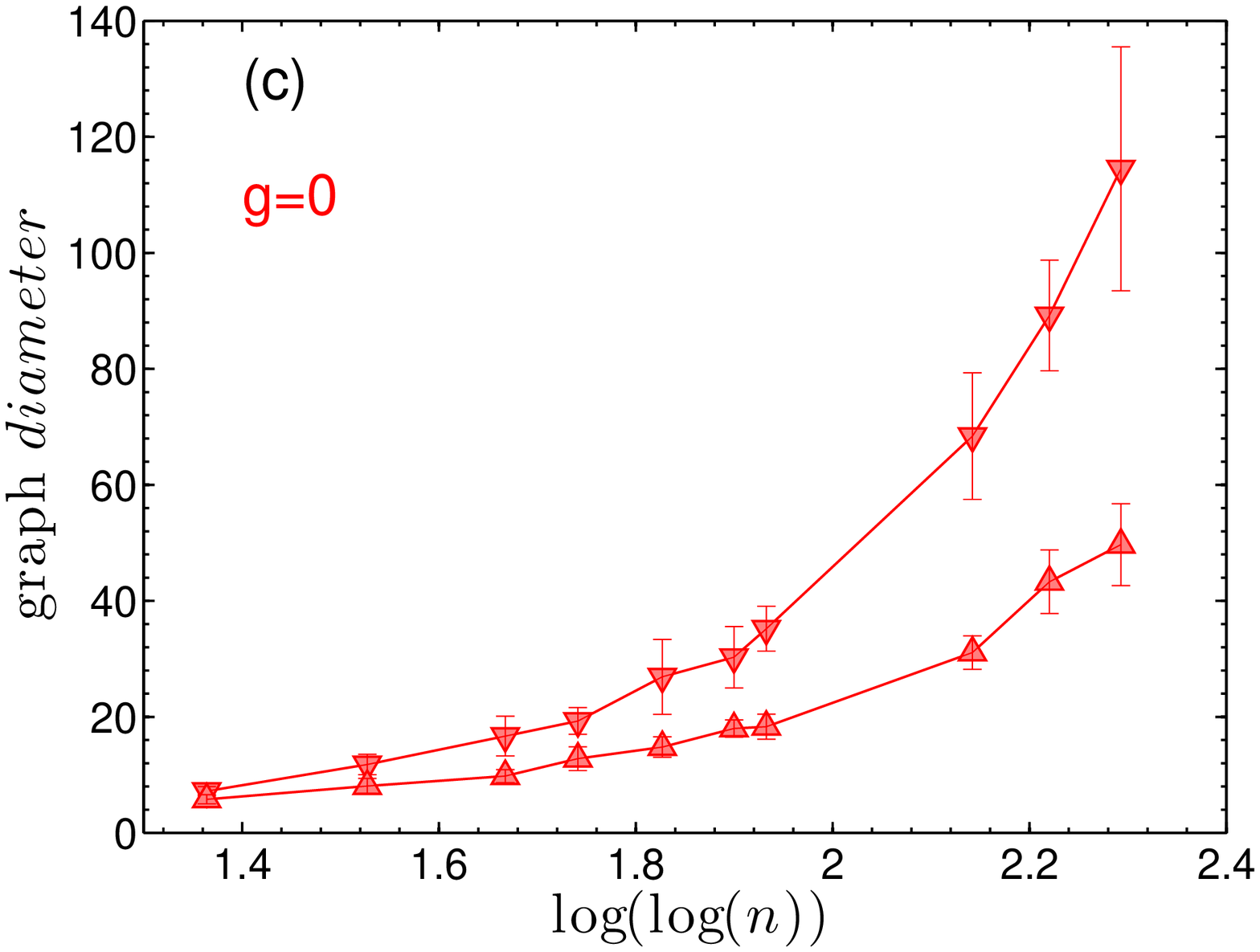}
&\includegraphics[width=0.5\columnwidth]{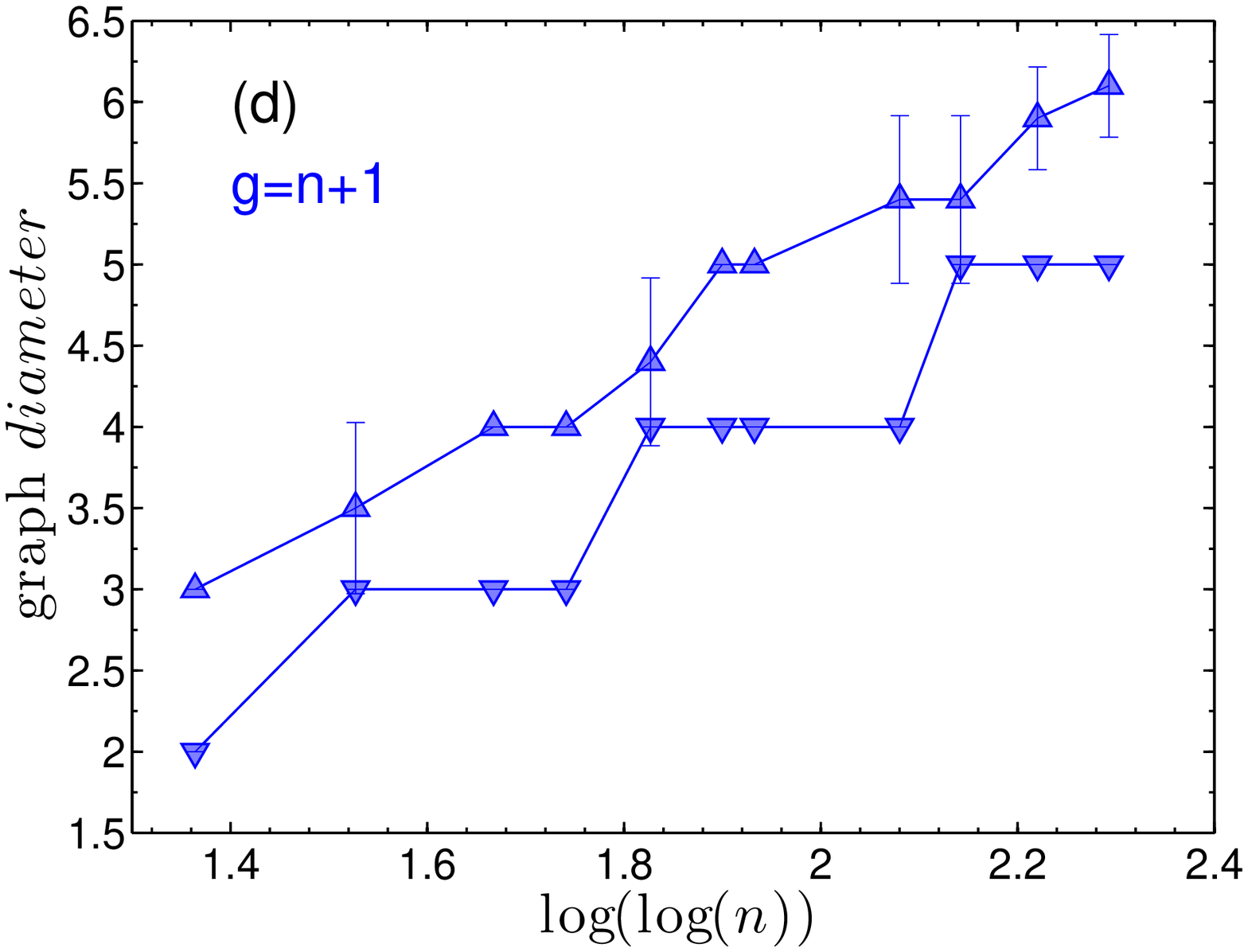}
\end{tabular}
\caption{\label{DiameterV1}
Variation of the graph-diameter with the number of vertices. 
Symbols: $\triangle$ random states;  $\triangledown$ frozen states.
The symbols correspond to averages over 10 independent constructions, the error bars are the standard deviations.
(a) Case with $g=0$ plotted vs. $log(n)$.
(b) Case with $g=n+1$ plotted vs. $log(n)$.
(c) Case with $g=0$ plotted vs. $log(log(n))$.
(d) Case with $g=n+1$ plotted vs. $log(log(n))$.
}
\end{figure}

\subsubsection{Scaling of the graph diameter with network size}
We have also investigated the effect of the surface genus on the scaling properties of maximal embedded graphs.
Networks can have different scaling properties which can be quantified by estimating the \emph{dimension} \cite{daqing2011dimension}.
For instance, a triangular lattice on a two-dimesional Euclidean surface has a  dimension equal to two.
This can be measured by investigating the scaling of the network-diameter  with the number of vertices in the network which, in this case, must scale with the square of the diameter and therefore we have the relation: ${diameter} \propto  n^{1/2}$. 
More generally, if the relation between the  diameter and the size is a power law, then one can define a  dimension $d$ from: $diameter \propto n^{1/d}$.

In Fig.\ref{DiameterV} the graph-diameter is plotted vs. the number of vertices for maximal embedded graphs with low genera ($g=0$) and high genera ($g=n+1$) both for random and frozen states.
As clearly visible from the plots, the genus strongly affects the value and the scaling of the diameters.

Incidentally, the dimension of maximal embedded graphs on spherical surfaces ($g=0$) has been studied in the context of two-dimesional quantum gravity where the (Euclidean) space-time is associated to a dynamical triangulation with all edge-lenghts constant but with variable vertex degree which is associated to the local space-time curvature \cite{Regge61,David85,Caracciolo88,Brugmann93}.
Studies in this context have revealed that the  dimension must be equal to 4 \cite{Agishtein90}; a result comforted by an analytical calculation via a transfer matrix formalism \cite{Kawai93}.
%When the temperature is lowered the complex  structure of the random triangulated surface is fattened down however it has been observed that it maintains a  dimension well above 2 \cite{Kownacki04}.
%To our knowledge, there are no previous studies of large triangulated surfaces with arbitrary genus.
%Graphs with small genus have larger diameter that scales almost linearly in log-log scale revealing therefore a power-law behaviour which is consistent with  dimensions $d= 2.2$ and   $d=2.8$ respectively for the  frozen and random states.
%The best fits for the  dimension for the high genus cases are respectively $d= 7.4$ and    $d= 8.8$ for the  frozen and random states.
In the present study we observe that the diameter follows rather well the functional form: $diameter \sim c_1 n^{1/4} + c_2$.
However, by looking carefully at Fig.\ref{DiameterV} one may note that the power law behaviour is satisfactory for $g=0$ but, instead, for high genus cases, the growth is slower than any power law.
This is made evident in Fig.\ref{DiameterV1} where the $diameter$ is plotted as function of $\ln(n)$ and $\ln(\ln(n))$. 
We can see clearly in Fig.\ref{DiameterV1}(a) that the diameter in low genus networks grows faster than $\ln(n)$ indicating a power law scaling.
On the other hand, we can see  from Fig.\ref{DiameterV1}(b) that the diameter in high genus networks does not grow any faster than $\ln(n)$ and the growth may even be consistent with a  linear increase with $\ln(\ln(n))$ suggesting that these networks could be ultra-small worlds.
Ultra-small world properties have been observed in scale-free networks \cite{Cohen03} but here -instead- this property is revealed also in the frozen state which are almost regular graphs with very narrow degree distributions.
It must be stressed that it is beyond the purpose of the present work to establish the exact scaling laws, indeed, for this purpose, a larger range of sizes must be explored.
However, what we have observed gives enough qualitative evidence to establish that the surface genus strongly affects the global properties of maximal embedded graphs changing the scaling of the diameter with the number of vertices passing from a large-world-kind of behaviour to a small--, and possible even ultrasmall--world behaviour. 
Very similar results are retrieved by considering the average distance between vertices instead of the diameter.

\begin{figure}
\centering
\includegraphics[width=0.8\columnwidth]{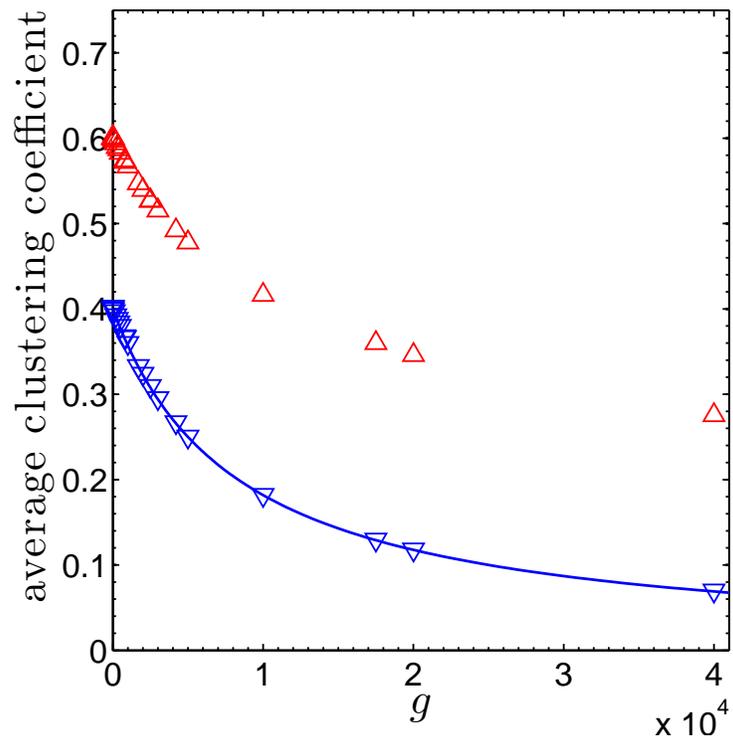}
\caption{\label{ClusteringCoeff}
Variation of the average clustering coefficient vs. the genus. 
Symbols: $\triangle$ random states;  $\triangledown$ frozen states.
The line is the prediction for the average clustering coefficient from $\bar c = 2(5+12(g-1)/n)^{-1}$ (see text).}
\end{figure}

\subsubsection{Clustering coefficient}

Another important and widely used graph-measure is the clustering coefficient which quantifies the level of local interrelations between a set of neighbouring vertices.
We measured the clustering coefficient of each vertex by taking the set of its neighbouring vertices and counting the number of edges between them and then dividing this number by the maximum possible number of edges between them (which is $k(k-1)/2$ for a vertex with degree $k$, i.e. all $k$ neighbors connected among themselves).
In Fig.\ref{ClusteringCoeff} we plot the average clustering coefficient for both random and frozen states.
We can observe that maximal embedded graphs have finite clustering coefficients which decrease with the increasing genus. 
This is related to the fact that the average number of neighbours increases with the genus (Eq.\ref{av_n}).
Indeed, in a triangulation, a vertex with degree $k$ has at least $k$ edges between its neighbours.
This means that, for a vertex with degree $k$ the clustering coefficient must have the following lower bond:
\begin{equation}\label{clust}
c(k) \ge \frac{2 k}{k(k-1)}=\frac{2 }{k-1}\;\;,
\end{equation}
which is a decreasing function of $k$.
The equality in Eq.\ref{clust} is achieved when the local structure surrounding the vertex with degree $k$ is a polyhedron with $k$ vertices. %, and this is the expected configuration for a regular 
If the degree distribution is narrowly distributed around $\left< k \right>$, then we can substitute $ \left< \frac{2 }{k-1}\right> \simeq  \frac{2 }{\left< k\right> -1}$ and therefore from Eqs. \ref{av_n} and \ref{clust} we can derive a lower bound on the average clustering coefficient:  $\bar c = \frac{2 }{\left< k\right> -1}=2(5+12(g-1)/n)^{-1}$ .
%random states have consistently a larger clustering coefficient than the frozen counterparts.
This lower bond is plotted with a continuous line in Fig.\ref{ClusteringCoeff}.
As one can see, the line overlaps well with the results for the frozen states.
This is intuitively correct because the frozen states are almost regular graphs with planar local configurations.

\section{Scale free reference network} \label{s.k*1}\label{s.powLaw}
Let us now consider a non-uniform reference degree distribution, and let us investigate the case where such a distribution is a power law: $p(k^*)\propto 1/(k^*)^{\alpha+1}$.
Networks with power law-kind degree distribution (scale free networks) are common to several natural and artificial systems and the study of their occurrence and properties is therefore of great interest  \cite{Barabasi99,Calrarelli07}.

\begin{figure}
\centering
%\begin{table}[cc]
\includegraphics[width=0.45\columnwidth]{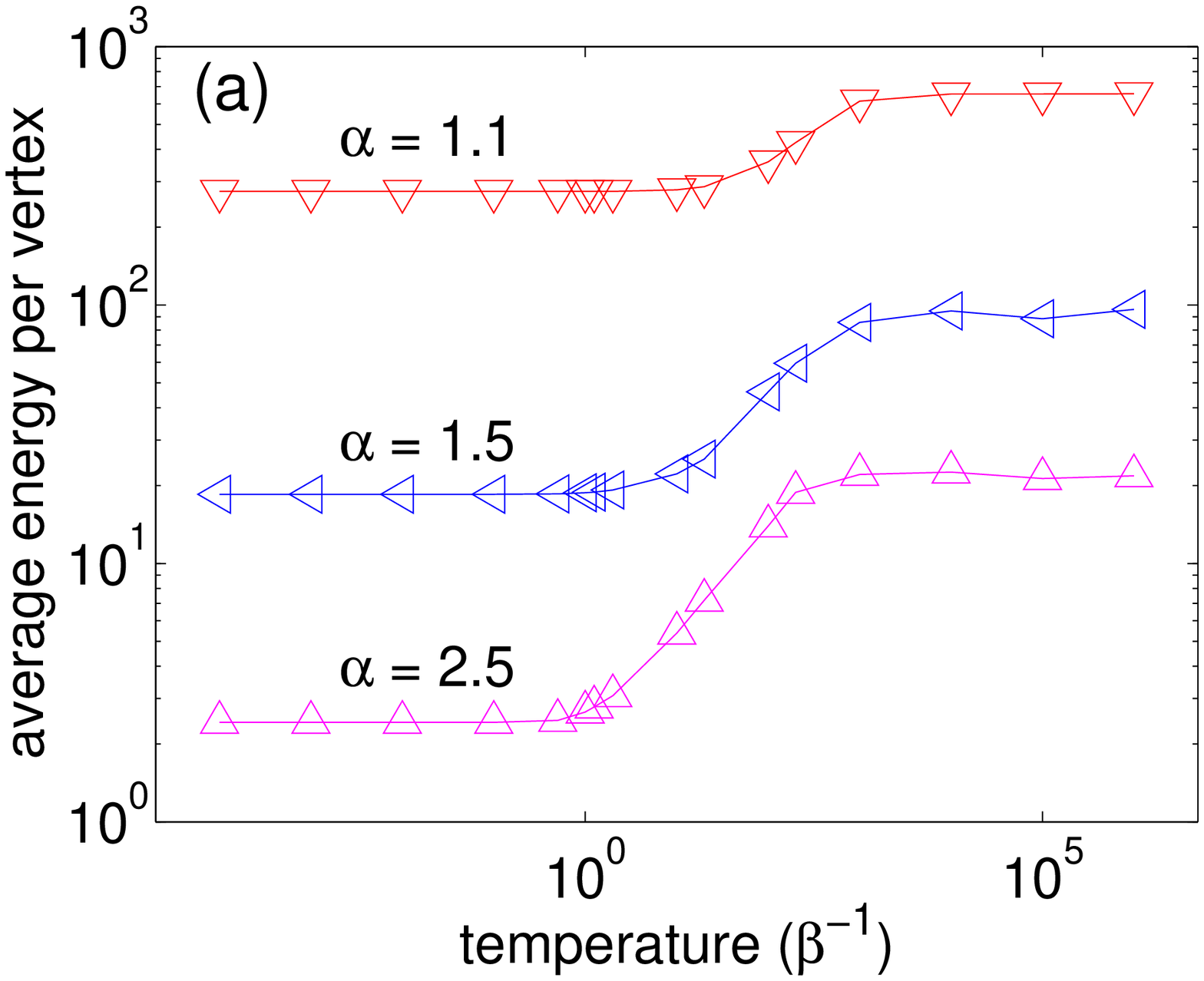}
\includegraphics[width=0.45\columnwidth]{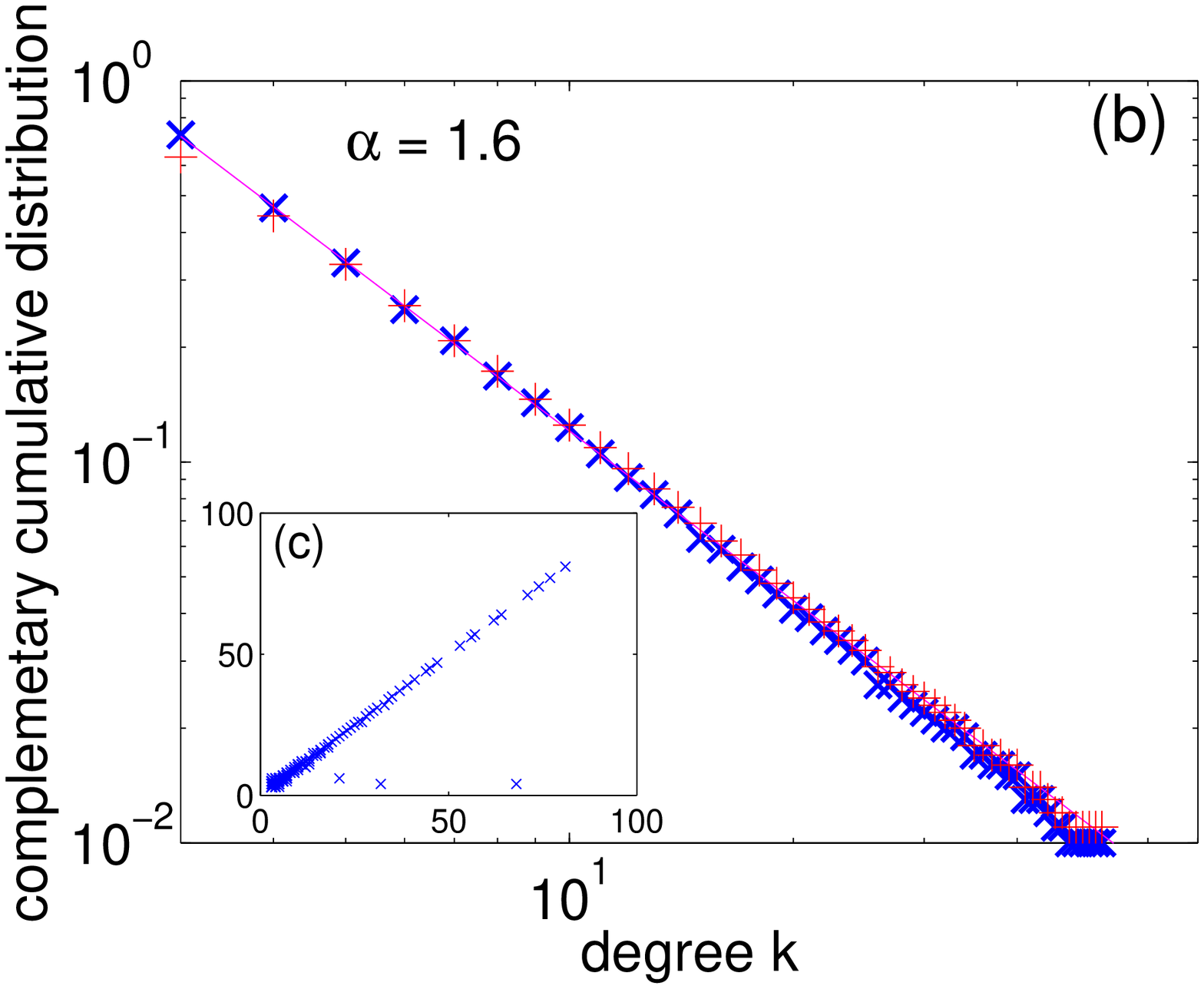} 
%\end{table}
\caption{\label{PowLawDegr}
(Color online) 
(a) Residual average energy per node at different temperatures for referential degree distributions with different values of the exponent $\alpha$. 
The maximally embedded graph has $n=1000$ vertices and it has been relaxed for $10^7$ steps with a Glauber--Kawasaki type of dynamics (Eq.\ref{prob}).
(b) Complementary cumulative distribution for an hyperbolic maximally embedded graph with $\alpha = 1.6$. %and $g=67$.
The red `+' symbols are the reference degree distribution $k^*$, whereas the blue '$\times$' symbols are the  degree distribution attained by the maximally embedded graph.
(c) The inset shows the degree of each vertex, $k_i$,   in the maximally embedded graph (y-axis) vs. the corresponding referential degree $k^*_i$ (x-axis).
}
\end{figure}

\begin{figure}
\centering
\includegraphics[width=0.45\columnwidth]{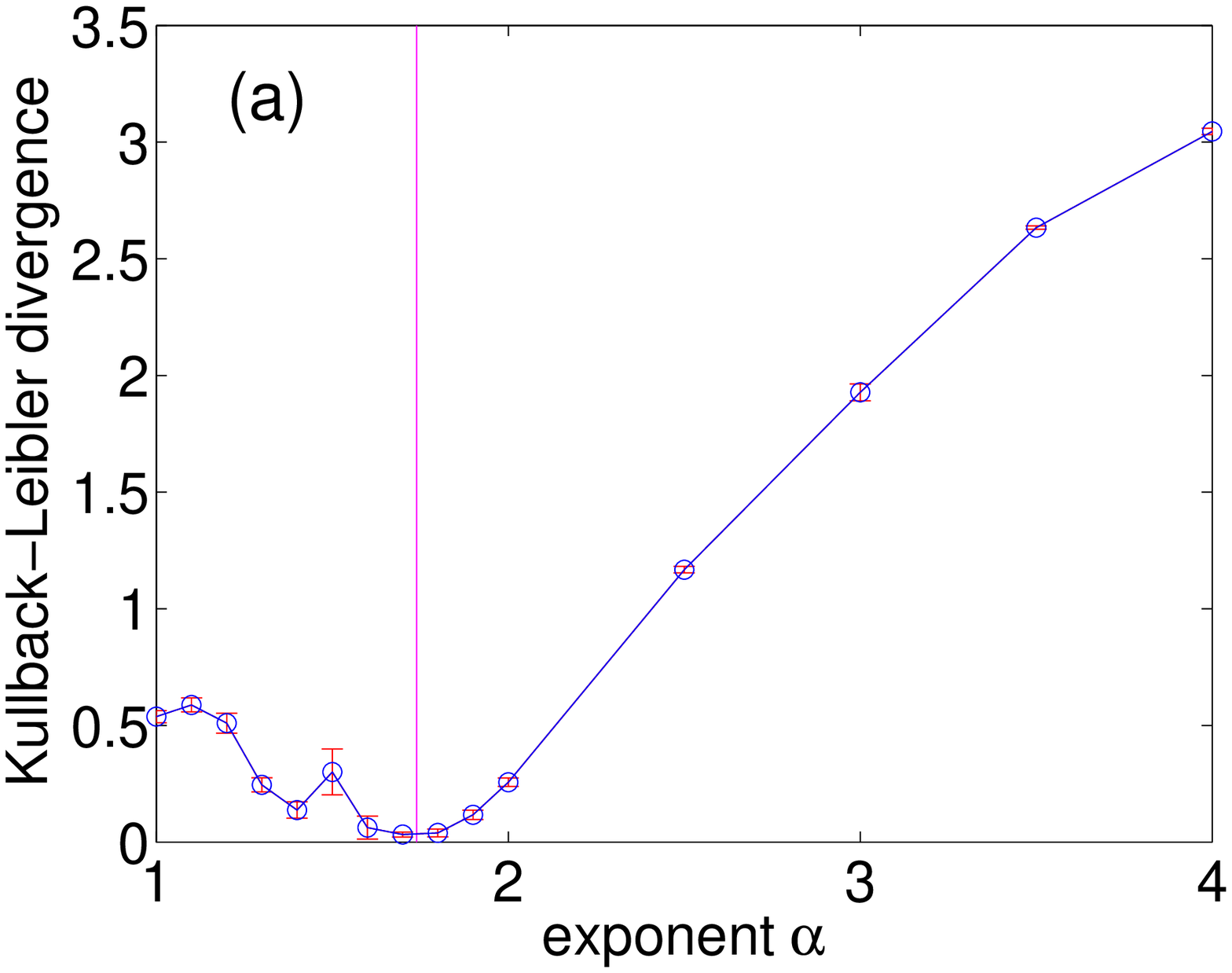} 
\includegraphics[width=0.45\columnwidth]{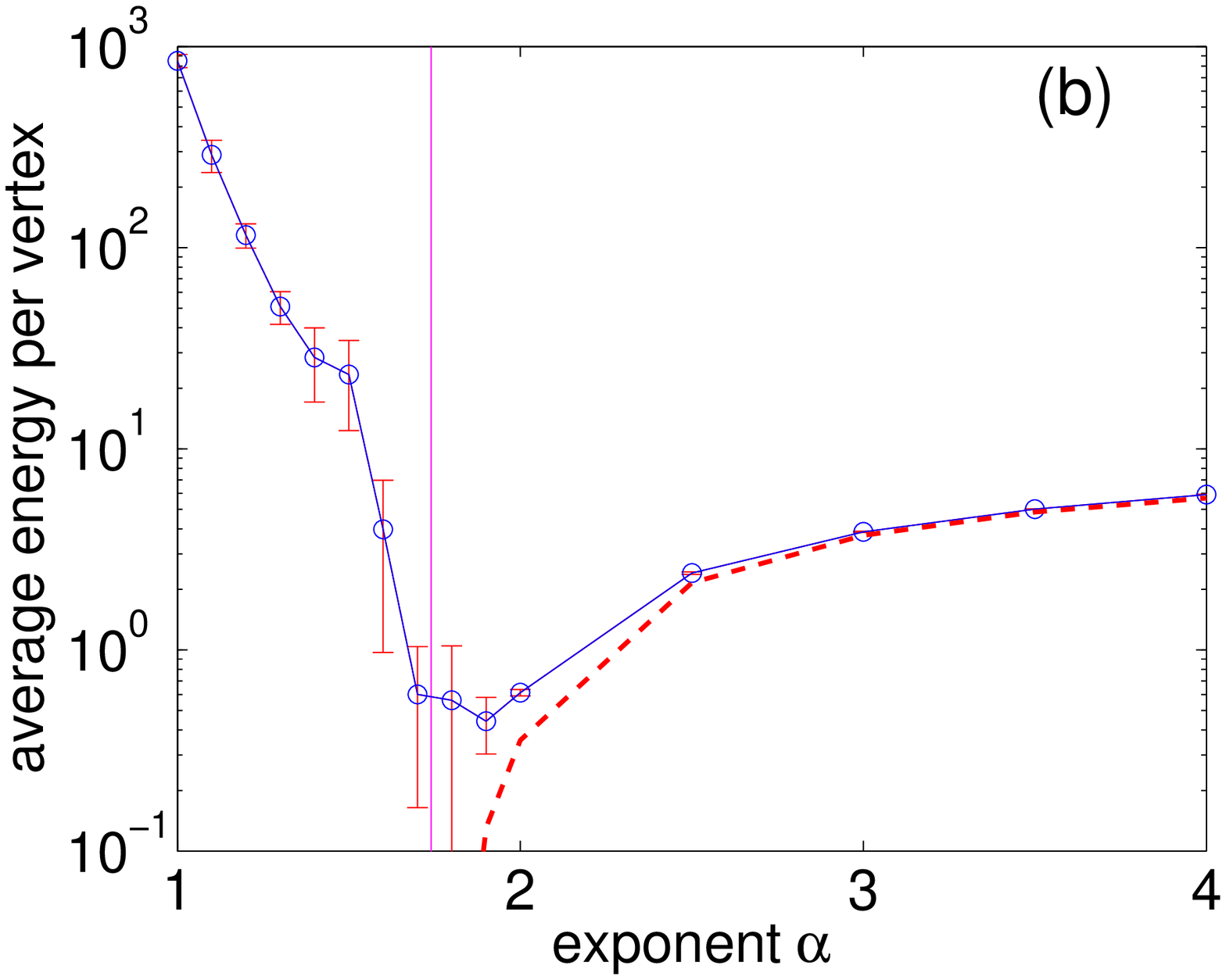} 
\caption{\label{PowLawDegrKLE}
(a) KullbackÐLeibler divergence between the reference degree distribution and the distribution of the embedded maximally embedded graph for various values of the exponent $\alpha$.
(b) Residual average energy per vertex, the dotted line is $(\left< k \right>-\left< k^* \right>)^2$.
The vertical lines indicate the point $\alpha = 1.74$ at which the reference distribution has $\left< k^* \right>=6$.
The symbols are averages over 10 simulations and the error bars are the standard deviations.
}
\end{figure}

%\subsection{Frozen states}
%It is known that planar triangulations can achieve power law degree distributions with different exponents  \cite{Andrade05,song2012}.
There are several known methods to build planar triangulations and planar graphs with power law degree distributions with different exponents  \cite{Andrade05,Zhou05maxPlanar,song2012}.
For instance, in the previous section we have shown that, at high genera, the degree distribution of the random state tends to broaden towards a power-law-like distribution. 
However, it has been so far unclear if, in general, \emph{any} power law degree distribution can be realized by using triangulated surfaces.
Here, we investigate this point by using an energy term \label{energyT1} with reference degree per each vertex, $k^*_i$, that follows a power law distribution $p(k^*)\propto 1/(k^*)^{\alpha+1}$. % with $3\le k^* \le n-1$.
In this case, the genus is set to a value that lead to an average degree which best approximates the reference average degree: 
$ \left< k \right>  = 6 + 12 \frac{g-1}{n} \simeq   \left< k^* \right>$ or conversely $g\simeq 1+ n (\left< k^* \right>-6)/12$ (see Eq.\ref{av_n}).
We can see that $ \left< k^* \right> = 6$ corresponds to $g=1$, instead $ \left< k^* \right> <6$ produces elliptic triangulations ($g=0$) and conversely  $ \left< k^* \right> > 6$ produces hyperbolic triangulations ($g>1$).

\subsection{Numerical simulations}
We generate a triangulation in a random state at high temperatures and then we quench it down to zero temperature.
These simulations have been carried over by using specifically developed numerical tools which are now made freely available from Matlab Central Ref.~\cite{HyperTriang}.
The random state has been already described in the previous section.
The quenching dynamics from random state to the frozen state is very similar to the one observed in the case of uniform reference distribution:  the system  dynamically slows down into a frozen state where the ground state is not reached in any finite time. 
Such a dynamical freezing is shown in Fig.\ref{PowLawDegr}(a)  where it is reported the average residual energy per vertex, after $10^4\times n$ time-steps with a Glauber--Kawasaki type of dynamics (Eq.\ref{prob}), at different temperatures $\beta^{-1}$ for triangulations with $n=1000$ vertices.
The frozen distribution for $\alpha = 1.6$ is shown in Fig.\ref{PowLawDegr}(b).
In this case $ \left< k \right>  =6.8$ and $g=67$. 
As one can see, the reference distribution (`+' symbols) is reproduced well by the embedded triangulation (`$\times$' symbols).
Furthermore,  in Fig.\ref{PowLawDegr}(c) it is shown that the degree of each vertex $k_i$ satisfactorily approaches the corresponding reference degree $k^*_i$ with a correlation coefficient between the two series above 99\%.

We  investigated also several different reference distributions across a range of power law exponents $\alpha$ measuring how well they can be approached by the triangulation. % through the Glauber--Kawasaki dynamics.
For this purpose we considered several power law  reference degree distributions with exponents in the range $1\le \alpha \le 4$ and we let the system to dynamically evolve with the Glauber--Kawasaki dynamics (Eq.\ref{prob}) towards a zero-temperature frozen limit.
Here we qualify the distance between the achieved distribution $p(k)$ and the reference distribution $p^*(k)$  by computing the Kullback--Leibler divergence: $KL = \sum_k p(k)\log_2 p(k)/p^*(k)$.
%Let here note that for a power law distribution, in the limit $n\rightarrow \infty$, one can calculate that the average degree is: $\left< k^* \right>=(\zeta(\alpha)-1-2^{-\alpha}))/(\zeta(\alpha+1)-1-2^{-\alpha-1})$ with $\zeta$ the Riemann zeta function.
%This is a decreasing function of $\alpha$ that diverges when $\alpha \rightarrow 1^+$ and becomes smaller than 6 for  $\alpha \sim 1.74$.
%This implies that the average degree of the embedded triangulation can match the average reference degree only within these two limits.
%However, power law distributions can also be obtained outside these bounds, with some deviations. 
In Fig.\ref{PowLawDegrKLE}(a) we report the values of the Kullback--Leibler divergence for various exponents $\alpha$. 
One can observe that  the two distributions have smallest differences in the region around $\alpha\sim 1.7$.
Let here note that for a power law distribution, in the limit $n\rightarrow \infty$, one can calculate analytically the average degree: $\left< k^* \right>=(\zeta(\alpha)-1-2^{-\alpha}))/(\zeta(\alpha+1)-1-2^{-\alpha-1})$ with $\zeta$ the Riemann zeta function.
This is a decreasing function of $\alpha$ that diverges when $\alpha \rightarrow 1^+$ and becomes equal to 6 for  $\alpha \sim 1.74$.
We have therefore that power laws degree distributions with  $\alpha < 1.74$ are associated with hyperbolic triangulations ($g>1$) whereas  $\alpha > 1.74$ are associated with planar or elliptic triangulations ($g=1$ or $0$).
It appears therefore that power law distributions are best retrieved for planar/elliptic triangulations.
Interestingly, around these values of the exponents $\alpha$ are associated to many known natural and artificial scale free networks \cite{Barabasi99,Calrarelli07}.
By looking at the final energy achieved (see Fig.\ref{PowLawDegrKLE}(b)) we can observe that for $\alpha > 2$ the residual energy per vertex almost coincides with $(\left< k \right>-\left< k^* \right>)^2$ indicating therefore that most of the differences in the distribution are consequence of the fact that, in this region of the exponents, the reference average degree becomes too small and it cannot be matched by the triangulation.

%\subsection{Freezing dynamics}
%We then performed Glauber--Kawasaki dynamics (Eqs.\ref{energyT1b}, \ref{prob}).

%We then tested whether, at zero temperatures, the reference  

\begin{figure}
\centering
\includegraphics[width=0.75\columnwidth]{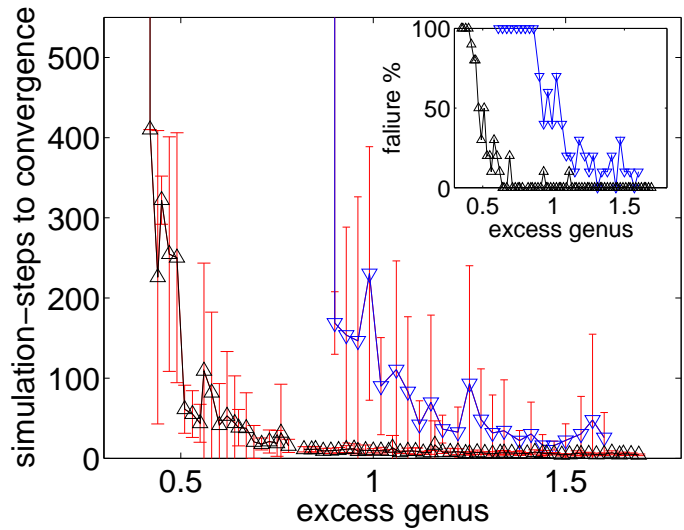} 
\caption{\label{EmbeddingConvergence}
(Color online) 
These plots illustrate that the convergence towards a maximally embedded graph that contains a given reference network as subgraph is always achieved when the genus is large enough.
The y-axis reports the number of steps to achieve the embedding in units of $n^2$ and the x-axis is the relative excess genus:  $(g-g^*)/n$  (where $g^* = 1+  \frac{\left< k^* \right> - 6}{12} n$ is the lower bound for the genus from Eq.\ref{low_g}).
The symbols are average values and the error bars are standard deviations over 10 runs.
The plots refer to $n=100$ case.
In the inset, the per-cent of failures to reach the embedding in less than $500\times n^2$ simulation steps is shown. 
The downward triangles `$\bigtriangledown$' refer to a scale free network with exponent $\alpha=1$; upper triangles `$\bigtriangleup$' refer to a scale free network with exponent $\alpha=2$. 
}
\end{figure}

\section{Embedding any network}\label{s.embed}
Let us now discuss the more general case when a reference network is given (any kind of complex network, not a triangulation) and a maximally embedded graph  containing this reference network as a subgraph is sought.
We know that this must be possible for a sufficiently high genus and Eq.\ref{low_g} gives an insight on the lower bound for $g$.
However, it is also known that the problem of finding the embedding for a given graph is NP-hard and finding the minimum genus of the surface on which the embedding can be build is NP-complete  (the graph-embedding problem) \cite{Thomassen89}.
Nonetheless, it is always possible to develop algorithms that can find sub-optimal embeddings in polynomial times.  
It is expected that, for such sub-optimal embedding, larger genus must correspond to faster convergence.
Here, we briefly discuss the case of embedding of scale free networks, searching for the lower values of the genus associated with convergence in polynomial time.
To this end, here we have developed a simple procedure that starts with a maximally embedded  network on a surface of genus $g$ containing  the same number of vertices $n$ as the reference network.   
In this procedure we first build  the network randomly and then we explore the possible embeddings with two moves: the T1 move, described previously, and a re-addressing move that changes the vertex indices trying to connect couples of vertices that are connected in the reference network.
The algorithm accepts moves that result in a maximally embedded graph with an increased number of connections among vertices that are connected in the reference network.
By using this simple procedure we observe that convergence is always achieved when the genus is sufficiently larger than the bound from Eq.\ref{low_g} ($g^* = 1+  \frac{\left< k^* \right> - 6}{12} n$).
On the other hand, when such a bound is approached, convergence becomes much slower and --eventually-- solution is not reached within the limits of the simulation time.
This is illustrated in Fig.\ref{EmbeddingConvergence} where the number of steps to reach the embedding (expressed in unit of $n^2$) are reported as function of the relative excess of genus $(g-g^*)/n$ for scale-free reference networks.
One can see that for large enough genera the embedding is always achieved in a time of the order of $O(n^2)$ but,  when the genus become smaller, the algorithm fails to find the embedding in $O(n^3)$ (see inset of Fig.\ref{EmbeddingConvergence} for the per-cents of failures to reach embedding within $500\times n^2$ simulation steps).
The transition between the phase where embedding is easily attained to the phase where no solutions are found is rather sharp and it depends on the network properties. 
For instance, Fig.\ref{EmbeddingConvergence}  shows the difference  between two cases for scale free networks with $\alpha=1$ and 2 respectively.
It is beyond the purpose of the present article to investigate these convergence issues that also depend on the algorithm used.
Our purpose here has been simply to prove, through examples, that given an arbitrary reference network, a maximally embedded graph containing the  reference network as sub-graph can be constructed. 
We verified that the  properties of the maximally embedded graph are strongly related to the properties of the reference network. 
For instance, the degree of each vertex in the maximally embedded graph and in the reference network are highly correlated.
This also implies that  Gauss-Bonnet space-curvature from the  maximally embedded graph is highly correlated, at local level, with the space-curvature required to naturally embed the reference network.
Consequently, by producing a maximally embedded graph which contains the reference network as a subgraph we also retrieve a \emph{geometrical} embedding of the reference network into an appropriate hyperbolic manifold.

\section{Conclusions}\label{s.3}
We have shown that by considering networks embedded on surfaces one can develop a powerful methodology that allows to study any kind of network within a unified approach.
The surface genus is a very important characteristic that strongly influences both local and global properties of the embedded graph. 
There are two simple, local elementary moves, T1 and T2, that  allow to explore the entire set of maximally embedded graph on a given surface.
By means of these moves one can develop a statistical mechanics approach which can be used to investigate networks with different degrees of disorder but with constrained complexity. 
Within such a statistical mechanics framework, we find that at high temperatures, the network is disordered and it reveals an exponential degree distribution when the surface genera is low; instead, at high genera, the equilibrium distribution becomes a power law  with faster than exponential cut-off.
At low temperatures, an energy function can be defined such that the network degree distribution tends toward a referential degree distribution.
However, we observe that the cooling  dynamics incurs in a slowing down and the ground state cannot be reached in a finite time. 
When the reference degree distribution is regular,  we observe that a mechanism of geometrical frustration, associated with fractional average degree, can reintroduce fast cooling dynamics showing that the occurrence of this glassy phase is related to the amount of disorder.
We show that a mean field model can describe correctly some of the network properties especially in the range of finite temperatures.

From a global structural perspective, the surface genus is constraining the degree of interwovenness of the network and we have showed that it is affecting the graph-diameter.
We found that networks developing on surfaces with small genera have large-world properties with the diameter scaling as a power law of the number of vertices; instead high genera surfaces lead to much smaller networks that might be small-world or even ultra-small-world.
Interestingly, this scaling is observed in networks prepared at high temperatures and possessing broad degree distributions as well as in networks cooled at low temperature with very regular degrees.  

This methodology is general and it can be used to study any network.
Indeed, we demonstrated that embedded triangulations can be build in such a way that they contain an arbitrary network as subgraph. 
These are sub-optimal embeddings which can be achieved in polynomial computational time when the surface-genus is large enough in comparison with the theoretical lower bound $g^*$ (Eq.\ref{low_g}).
We have pointed out that these topological embeddings onto  maximally embedded graphs directly lead to geometrical embedding into spaces with non-Euclidean curvature (typically hyperbolic for complex reference networks).
Let us note that our approach is however much richer than a simple geometrical embedding onto the Poincarr\'e disk, indeed the surface genus also provides  `wormholes'  through which the embedding space is offering ways for shortcuts between otherwise distant points.
This further contraction of the embedding space is a key element that leads to ultra-small networks. 
The resulting geometrical embedding is on a complex hyperbolic manifold.

Future studies will focus on the application of these networks to information filtering by constructing maximal embedded graphs from a similarity measure in analogy with what already done in the case of planar maximally embedded graphs with the PMFG construction  \cite{Tumminello05,song2012hierarchical,song2011nested} that in the present context corresponds to the $g=0$ case.

\section*{Acknowledgement}
This work was partially supported by COST MP0801 project.\\

\bibliographystyle{unsrt}
%\bibliography{/Users/tomaso/TOM/Biblio/GeneralBiblio.bib}

\begin{thebibliography}{10}

\bibitem{Barabasi99}
A.L. Barab\'asi and R.~Albert.
\newblock Emergence of scaling in random networks.
\newblock {\em Nature}, 286:509--512, 1999.

\bibitem{Newman03}
M.~E.~J. Newman.
\newblock The structure and function of complex networks.
\newblock {\em SIAM REVIEW}, 45:167--256, 2003.

\bibitem{zhou2006}
T.~Zhou, B.H. Wang, PM~Hui, and KP~Chan.
\newblock Topological properties of integer networks.
\newblock {\em Physica A: Statistical Mechanics and its Applications},
  367:613--618, 2006.

\bibitem{Calrarelli07}
Guido Caldarelli.
\newblock {\em Scale-Free Networks: Complex Webs in Nature and Technology}.
\newblock Oxford Univesity Press, 2007.

\bibitem{cohen2010complex}
R.~Cohen and S.~Havlin.
\newblock {\em Complex Networks: Structure, Robustness and Function}.
\newblock Cambridge Univ Pr, 2010.

\bibitem{Barth04}
W.~Barth.
\newblock {\em Compact complex surfaces}, volume~4.
\newblock Springer Verlag, 2004.

\bibitem{song2011nested}
W.-M. Song, T.~Di~Matteo, and T.~Aste.
\newblock Nested hierarchies in planar graphs.
\newblock {\em Discrete Applied Mathematics}, 159:2135--2146, 2011.

\bibitem{AsteSherr11}
Tomaso Aste, Ruggero Gramatica, and T.~Di~Matteo.
\newblock Random and frozen states in complex triangulations.
\newblock {\em Philosophical Magazine}, 92:244--254, 2012.

\bibitem{Andrade05}
J.~S. Andrade~Jr., H.~J. Herrmann, R.~F.~S. Andrade, and L.~R. da~Silva.
\newblock Apollonian networks: Simultaneously scale-free, small world,
  euclidean, space-filling and with matching graphs.
\newblock {\em Phys. Rev. Lett.}, 94:018702, 2005.
\newblock e-print: cond-mat/0406295.

\bibitem{Zhou05maxPlanar}
Tao Zhou, Gang Yan, and Bing-Hong Wang.
\newblock Maximal planar networks with large clustering coefficient and
  power-law degree distribution.
\newblock {\em Phys. Rev. E}, 71:046141, Apr 2005.

\bibitem{gu2005simplex}
Z.M. Gu, T.~Zhou, B.H. Wang, G.~Yan, C.P. Zhu, and Z.Q. Fu.
\newblock Simplex triangulation induced scale-free networks.
\newblock {\em Dynamics of Continuous, Discrete and Impulsive Systems B},
  13:505--510, 2006.

\bibitem{song2012}
W.-M. Song, T.~Di~Matteo, and T.~Aste.
\newblock Building complex networks with platonic solids.
\newblock {\em Phys. Rev. E}, 85(4):046115, 2012.

\bibitem{Ringel74}
G.~Ringel.
\newblock {\em Map Color Theorem}.
\newblock Springer-Verlag, Berlin, 1974.

\bibitem{Alexander30}
J.~W. Alexander.
\newblock The combinatorial theory of complexes.
\newblock {\em Ann. of Math.}, 31:294--322, 1930.

\bibitem{aste2Dfroth}
H.M. Ohlenbusch, T.~Aste, B.~Dubertret, and N.~Rivier.
\newblock The topological structure of 2d disordered cellular systems.
\newblock {\em Eur. Phys. J. B}, 29:211--220, 1998.

\bibitem{Dubertret98}
B.~Dubertret, T.~Aste, H.~M. Ohlenbusch, and N.~Rivier.
\newblock Two-dimensional froths and the dynamics of biological tissues.
\newblock {\em Phys. Rev. E}, 58(5):6368--6378, Nov 1998.

\bibitem{HyperTriang}
T.~Aste
\newblock A toolbox to generate triangulations on surfaces with arbitrary genera
\newblock {\em Matlab Central: http://www.mathworks.com/matlabcentral/fileexchange/37912}

\bibitem{ADHhypnet04}
T.~Aste, T.~Di{~}Matteo, and S.T. Hyde.
\newblock Complex networks on hyperbolic surfaces.
\newblock {\em Physica A}, 346:20--26, 2005.

\bibitem{boguna2010sustaining}
M.~Bogu{\~n}{\'a}, F.~Papadopoulos, and D.~Krioukov.
\newblock Sustaining the internet with hyperbolic mapping.
\newblock {\em Nature Communications}, 1:62, 2010.

\bibitem{Krioukov10}
Dmitri Krioukov, Fragkiskos Papadopoulos, Maksim Kitsak, Amin Vahdat, and
  Mari\'an Bogu\~n\'a.
\newblock Hyperbolic geometry of complex networks.
\newblock {\em Phys. Rev. E}, 82:036106, Sep 2010.

\bibitem{barthelemy2011spatial}
M.~Barth{\'e}lemy.
\newblock Spatial networks.
\newblock {\em Physics Reports}, 499(1):1--101, 2011.

\bibitem{daqing2011dimension}
L.~Daqing, K.~Kosmidis, A.~Bunde, and S.~Havlin.
\newblock Dimension of spatially embedded networks.
\newblock {\em Nature Physics}, 2011.

\bibitem{misner1973gravitation}
C.W. Misner, K.S. Thorne, and J.A. Wheeler.
\newblock {\em Gravitation}.
\newblock WH Freeman \& co, 1973.

\bibitem{Richeson08}
D.S. Richeson.
\newblock {\em Euler's gem: the polyhedron formula and the birth of topology}.
\newblock Princeton Univ Pr, 2008.

\bibitem{RingelPNAS68}
G.~Ringel and JWT Youngs.
\newblock Solution of the heawood map-coloring problem.
\newblock {\em Proceedings of the National Academy of Sciences of the United
  States of America}, 60(2):438, 1968.

\bibitem{Kuratowski30}
K.~Kuratowski.
\newblock Sur le probleme des courbes gauches en topologie.
\newblock {\em Fund. Math}, 15(271-283):79, 1930.

\bibitem{AsteSherr}
T.~Aste and D.~Sherrington.
\newblock Glass transition in self organizing cellular patterns.
\newblock {\em J. Phys. A: Math. Gen.}, 32:7049--56, 1999.

\bibitem{Davison00}
Lexie Davison and David Sherrington.
\newblock Glassy behaviour in a simple topological model.
\newblock {\em Journal of Physics A: Mathematical and General}, 33(48):8615,
  2000.

\bibitem{Sherrington02}
David Sherrington, Lexie Davison, Arnaud Buhot, and Juan~P Garrahan.
\newblock Glassy behaviour in simple kinetically constrained models:
  topological networks, lattice analogues and annihilation-diffusion.
\newblock {\em Journal of Physics: Condensed Matter}, 14(7):1673, 2002.

\bibitem{Kownacki04}
{J.-P. Kownacki}.
\newblock Freezing of triangulations.
\newblock {\em Eur. Phys. J. B}, 38(3):485--494, 2004.

\bibitem{Eckmann07}
Jean-Pierre Eckmann.
\newblock A topological glass.
\newblock {\em J Stat Phys}, 129:289--309, 2007.

\bibitem{eckmann2012decay}
J.P. Eckmann and M.~Younan.
\newblock Decay of correlations in a topological glass.
\newblock {\em Philosophical Magazine}, 92(1-3):98--119, 2012.

\bibitem{Tutte62}
W.~Tutte.
\newblock A census of planar triangulations.
\newblock {\em Canadian J. Math.}, 14:21--38, 1962.

\bibitem{Brezin1978}
E.~Brezin, C.~Itzykson, G.~Parisi, and J.~B. Zuber.
\newblock {Planar Diagrams}.
\newblock {\em Commun. Math. Phys.}, 59:35, 1978.

\bibitem{boulatov86}
D.~V. Boulatov, V.~A. Kazakov, I.~K. Kostov, and A.~A. Migdal.
\newblock {Analytical and numerical study of a model of dynamically
  triangulated random surfaces}.
\newblock {\em Nucl. Phys. B}, 275:641, 1986.

\bibitem{Brezin90}
E.~Brezin and V.~A. Kazakov.
\newblock Exactly solvable field theories of closed strings.
\newblock {\em Phys. Lett.}, B236:144--150, 1990.

\bibitem{Gross90}
David~J. Gross and Alexander~A. Migdal.
\newblock {Nonperturbative Two-Dimensional Quantum Gravity}.
\newblock {\em Phys. Rev. Lett.}, 64:127, 1990.

\bibitem{Godreche92}
C.~Godr{\`e}che, I.~Kostov, and I.~Yekutieli.
\newblock Topological correlations in cellular structures and planar graph
  theory.
\newblock {\em Phys. Rev. Lett.}, 69(18):2674--2677, 1992.

\bibitem{angell00}
C.~A. Angell, K.~L. Ngai, G.~B. McKenna, P.~F. McMillan, and S.~W. Martin.
\newblock Relaxation in glassforming liquids and amorphous solids.
\newblock {\em Journal of Applied Physics}, 88(6):3113--3157, 2000.

\bibitem{Cavagna09}
Andrea Cavagna.
\newblock Supercooled liquids for pedestrians.
\newblock {\em Physics Reports-review Section of Physics Letters}, 476:51--124,
  2009.

\bibitem{Regge61}
T.~Regge.
\newblock General relativity without coordinates.
\newblock {\em Il Nuovo Cimento Series 10}, 19:558--571, 1961.

\bibitem{David85}
F.~David.
\newblock Planar diagrams, two-dimensional lattice gravity and surface models.
\newblock {\em Nuclear Physics B}, 257:45 -- 58, 1985.

\bibitem{Caracciolo88}
Sergio Caracciolo and Andrea Pelissetto.
\newblock Nonperturbative lattice gravity.
\newblock {\em Nuclear Physics B - Proceedings Supplements}, 4:78 -- 82, 1988.

\bibitem{Brugmann93}
Bernd Br\"ugmann and Enzo Marinari.
\newblock 4d simplicial quantum gravity with a nontrivial measure.
\newblock {\em Phys. Rev. Lett.}, 70(13):1908--1911, Mar 1993.

\bibitem{Agishtein90}
M.~E. Agishtein and A.~A. Migdal.
\newblock 1990 int. j. mod. phys. c.
\newblock {\em Int. J. Mod. Phys. C}, 165:1, 1990.

\bibitem{Kawai93}
H.~Kawai, N.~Kawamoto, T.~Mogami, and Y.~Watabiki.
\newblock Transfer matrix formalism for two-dimensional quantum gravity and
  fractal structures of space-time.
\newblock {\em Physics Letters B}, 306:19--26, 1993.

\bibitem{Cohen03}
Reuven Cohen and Shlomo Havlin.
\newblock Scale-free networks are ultrasmall.
\newblock {\em Phys. Rev. Lett.}, 90(5):058701, Feb 2003.

\bibitem{Thomassen89}
Carsten Thomassen.
\newblock The graph genus problem is np-complete.
\newblock {\em Journal of Algorithms}, 10(4):568 -- 576, 1989.

\bibitem{Tumminello05}
M.~Tumminello, T.~Aste, T.~Di{~}Matteo, and R.N. Mantegna.
\newblock A tool for filtering information in complex systems.
\newblock {\em Proc. Natl. Acad. Sci.}, 102:10421--10426, 2005.

\bibitem{song2012hierarchical}
W.-M. Song, T.~Di~Matteo, and T.~Aste.
\newblock Hierarchical information clustering by means of topologically
  embedded graphs.
\newblock {\em PLoS ONE}, 7:e31929, 2012.

\end{thebibliography}

\end{document}